\documentclass[apj]{emulateapj}
\usepackage{natbib}
\usepackage{amssymb}
\usepackage{amsmath}
\def\be{\begin{eqnarray}}
\def\ee{\end{eqnarray}}
\usepackage{comment}
\usepackage{graphicx,subfigure}
\usepackage{enumerate}
\newcommand{\pasa}{{PASA}}
\usepackage{tgtermes}

\shorttitle{VLA follow-up of gravitational waves during aLIGO O1}

\shortauthors{Palliyaguru, Corsi, Kasliwal et al.}

\begin{document}

\title{Radio follow-up of gravitational wave triggers during Advanced LIGO O1}
\author{N.~T.~Palliyaguru\altaffilmark{1}, A.~Corsi\altaffilmark{1}, M.~M.~Kasliwal\altaffilmark{2}, S.~B.~Cenko\altaffilmark{3,4}, D.~A.~Frail\altaffilmark{5}, D.~A.~Perley\altaffilmark{6}, N.~Mishra\altaffilmark{1,7}, L.~P.\,Singer\altaffilmark{3}, A.~Gal-Yam\altaffilmark{8}, P.~E.~Nugent\altaffilmark{9,10}, J.~A.~Surace\altaffilmark{11}}
\altaffiltext{1}{Department of Physics, Texas Tech University, Box 41051, Lubbock, TX 79409, USA. E-mail: alessandra.corsi@ttu.edu}
\altaffiltext{2}{Division of Physics, Mathematics, and Astronomy, California Institute of Technology, Pasadena, CA 91125, USA.}
\altaffiltext{3}{Astrophysics Science Division, NASA Goddard Space Flight Center, Mail Code 661, Greenbelt, MD 20771, USA.}
\altaffiltext{4}{Joint Space-Science Institute, University of Maryland, College Park, MD 20742, USA.}
\altaffiltext{5}{National Radio Astronomy Observatory, P.O. Box O, Socorro, NM 87801, USA}
\altaffiltext{6}{Dark Cosmology Centre, Niels Bohr Institute, University of Copenhagen,  Juliane  Maries  Vej  30,  2100  Copenhagen,  Denmark.}
\altaffiltext{7}{Westview High School, 4200 NW 185th Ave, Portland, OR 97229, USA.}
\altaffiltext{8}{Benoziyo  Center  for  Astrophysics,  Weizmann  Institute  of Science, 76100 Rehovot, Israel.}
\altaffiltext{9}{Astronomy Department, University of California at Berkeley, Berkeley, CA 94720, USA.}
\altaffiltext{10}{Lawrence Berkeley National Laboratory, 1 Cyclotron Road MS 50B-4206, Berkeley, CA 94720, USA.}
\altaffiltext{11}{Spitzer Science Center, MS 220-6, California Institute of Technology, Pasadena, CA 91125, USA.}

\begin{abstract}
We present radio follow-up observations carried out with the Karl G. Jansky Very Large Array during the first observing run (O1) of the Advanced Laser Interferometer Gravitational-wave Observatory (LIGO). A total of three gravitational wave triggers were followed up during the $\approx 4$ months of O1, from September 2015 to January 2016. Two of these triggers, GW150914 and GW151226, are binary black hole merger events of high significance. A third trigger, G194575, was subsequently declared as an event of no interest (i.e., a false alarm). Our observations targeted selected optical transients identified by the intermediate Palomar Transient Factory (iPTF) in the Advanced LIGO error regions of the three triggers, and a limited region of the gravitational wave localization area of G194575 not accessible to optical telescopes due to Sun constraints, where a possible high-energy transient was identified. No plausible radio counterparts to GW150914 and GW151226 were found, in agreement with expectations for binary black hole mergers. We show that combining optical and radio observations is key to identifying contaminating radio sources that may be found in the follow-up of gravitational wave triggers, such as emission associated to star formation and AGN. We discuss our results in the context of the theoretical predictions for radio counterparts to gravitational wave transients, and describe our future plans for the radio follow-up of Advanced LIGO (and Virgo) triggers.
\end{abstract}


\section{Introduction}
\label{intro}
The first observing run (O1) of the Advanced Laser Interferometer Gravitational--wave Observatory \citep[LIGO;][]{abbot2016a}
started in September 2015 and ended in January 2016.
On 14 September 2015, the two Advanced LIGO detectors recorded their first significant event, GW150914, produced by the merger of two black holes (BHs) with masses of $\approx 36$\,M$_\odot$ and $\approx 29$\,M$_\odot$ \citep{abbot2016b}.  A panchromatic (radio to $\gamma$-rays) follow-up campaign of GW150914 was carried out by partner electromagnetic (EM) facilities, marking the start of GW astronomy \citep[][and references therein]{abbot2016c}. No high-significance EM counterpart to GW150914 was found, although a $\gamma$-ray event (compatible with the large GW localization area) occurring $\approx 0.4$\,s after GW150914 was reported by the \textit{Fermi}/GBM team \citep{cbg16}.  
A second GW alert was sent out to EM partners in October 2015, but the event that triggered this alert (G194575) was subsequently retracted as a false alarm \citep{gcn18626}. 
Finally, in December 2015, Advanced LIGO recorded a second significant binary BH event, GW151226 \citep{abbot2016d}. This detection was also accompanied by a vast EM follow-up effort \citep{aaa16,cbs16,csp16,ekp16,ggh16,rbg16,scs16}. 

Stellar mass BH binaries such as the ones that produced GW150914 and GW151226 are generally not expected to be associated with detectable EM signatures \citep[see e.g.,][]{mpr10}. However, after the first LIGO detection was announced, and following the \textit{Fermi/}GBM claim, several ideas have emerged about the possibility of producing EM counterparts to events like GW150914.
Proposed models range from EM signals emitted from in-falling matter from a circum-binary disk, to dissipation of magnetic energy and extraction of rotational energy from a BH via the Blandford-Znajek process \citep{bz77}.  Fallback material from a supernova (SN) explosion may remain bound around a BH for long periods and re-ignite accretion as the binary tightens \citep{plg16}. 
The core-collapse of a massive star may lead to the formation of ``blobs'' that evolve separately and merge, with some remaining core material falling into the merger remnant to produce a $\gamma$-ray burst \citep[GRB;][]{l16}, although this scenario is controversial \citep{wb16}. It has also been suggested that BH merger events of sufficient energy ($\sim 10^{49}$\,erg in collimation--corrected kinetic energy) occurring in dense enough circum-stellar material (CSM), and/or powering fast outflows that shock the ISM, may produce radio signatures of a few to a few tens of $\mu$Jy at 1.4 GHz \citep{yao16,Murase2016}.

The existence of detectable EM emission from binary BH mergers remains speculative. However, Advanced LIGO should detect 0.4--400 Neutron Star (NS)-NS and 0.2--300 NS-BH binaries per year when reaching nominal sensitivity \citep{abadie2010}. Compact binary mergers involving at least one NS are expected to produce EM emission. Indeed, these events could power short GRBs \citep[see e.g.][ and references therein]{b14}, ``kilonova'' emission in the optical \citep[e.g.,][]{lp98,mb12,bk13,mf14,jin16}, and detectable radio signatures \citep[as the ultra-to-mildly relativistic components of the ejecta interact with the ISM; e.g.,][]{np11,hnh16}. Optical kilonova and radio emissions, unlike the $\gamma-$rays, are less affected by collimation, relativistic beaming, and viewing angle effects, thus offering a better chance of detection for binary NS (or BH-NS) mergers launching relativistic jets not aligned with our line of sight (which are expected to constitute the majority of Advanced LIGO detections). 

Given the large error areas associated with Advanced LIGO localizations \citep{cbb06,f11,kvd11,nsd11,abbot2016e}, the detection of an EM counterpart via large field-of-view (FOV) facilities such as the intermediate Palomar Transient Factory \citep[iPTF;][]{lkd09} can provide orders of magnitude better localizations, and enable panchromatic follow-up observations with smaller FOV instruments like the Karl G. Jansky Very Large Array\footnote{The National Radio Astronomy Observatory is a facility of the National Science Foundation operated under cooperative agreement by Associated Universities, Inc. } \citep[VLA;][]{pnj09}.  In turn, a radio detection in coincidence with one of the many optical candidates expected to be found in the error-area of a GW event can help enhance the confidence in the EM counterpart association, rule out false associations, and constrain source and environment properties. While GWs from a tidally disrupted NS in a NS--BH binary can provide better constraints on the equation of state  \citep[via measurements of the NS radius;][]{v00}, radio follow-up can provide unique information such as outflow velocity, geometry and total energetics \citep[through calorimetry;][]{fwk00}, and circum-binary medium density \citep{b10,hnh16}. Our O1 VLA follow-up program\footnote{VLA/15A-339; PI: Corsi} was designed to achieve these goals (while leaving space for serendipitous discoveries).

Here, we present the results of our VLA follow-up observations of GW150914, G194575, GW151226, carried out  in coordination with the iPTF. We refer the reader to \citet{kcs16} and \citet{k16} for details regarding the iPTF follow-up strategy, selection criteria, and photometric/spectroscopic properties of candidate optical counterparts found in the error regions of the LIGO triggers. Here, we focus on the results of our VLA observations. Specifically, in  Section~2 we describe our VLA follow--up program for O1, and the criteria based on which we identified transients worthy of radio follow-up. In Section~3 we describe our radio observations and image processing.
In Sections~4-5 we discuss our results. Finally, in Section~6, we give an overview of future prospects for our VLA follow-up program and conclude. 

\begin{table*}
\label{tb:trigger_info}
\begin{center}\begin{footnotesize}
\caption{GW trigger ID, GW trigger time, GW signal-to-noise ratio (SNR), redshift range (from the GW observations), and nature (type) of the three GW triggers that were sent to EM observers during Advanced LIGO O1. For each GW trigger, we also indicate the 90\% GW localization area \citep{abbot2016f}, the area that was imaged by the iPTF, the containment probability (i.e., the probability that the iPTF imaged area contains the true location of the source), the number of optical transients identified by the iPTF in such area, and the number of candidates followed-up via our VLA program. References are as follows: [1] \citet{gcn18420}; [2] \citet{gcn18474}; [3] \citet{gcn18337}; [4] \citet{abbot2016b}; [5] \citet{kcs16}; [6] \citet{gcn18914}; [7] \citet{gcn18442}; [8] \citet{gcn18528}; [9] \citet{gcn18560}; [10] \citet{gcn18584}; [11] \citet{gcn18497}; [12] \citet{gcn18762}; [13] \citet{abbot2016d};  [14] \citet{gcn18846}; [15] \citet{gcn18780}; [16] \citet{gcn18873}.  \label{tb:trigger_info}
}
\begin{tabular}{lllllllllll}
\hline
ID & GW trigger time  &SNR& $z$ & Type & 90\% GW area & iPTF area & Cont. prob. & no. iPTF cand. & no. VLA cand. & Ref.\\ 
     &     (MJD)              &    &        &       &  (deg$^2$)       & (deg$^2$)         &                            &       \\
\hline
GW150914& 57279.410 &24 & $0.054-0.136$& BH--BH&230 & 126 & 2.5\% & 8 &1 & [1-6]\\
G194575& 57317.566 &--& --& --& -- & 1114&--& 42 & 13 &[7-11]\\
GW151226& 57382.166 &13& $0.05-0.13$& BH--BH& 850 & 952 & 51\% & 20 & 2&[12-15]\\
\hline
\end{tabular}
\end{footnotesize}\end{center}\end{table*}

\section{The VLA follow--up program}
\label{VLA_program}
Given the relatively small VLA FOV (primary beam) of a few arcmin at  a few GHz frequencies  \citep[to be compared to {$\mathcal O$}(100) deg$^2$ GW localization areas; ][]{abbot2016e}, 
our VLA follow-up program of Advanced LIGO triggers was designed to work in coordination with larger FOV facilities, such as the iPTF \citep{kcs16}. The iPTF large FOV, as well as its automated transient identification and selection capabilities, assure that candidate EM counterparts worthy of follow-up are promptly identified. 

Some of the advantages of following--up in the radio selected optical candidates identified in the LIGO localization regions (as opposed to a ``blind'' coverage of the GW localization area in the radio), are the fact that optical (photometric and spectroscopic) observations can provide: 
\begin{itemize}
\item Accurate source locations, that enable deeper sensitivity radio follow-ups with small FOV radio facilities; 
\item Redshift measurements, that can be compared for consistency with the distance estimate derived from the GW triggers; 
\item Host galaxy information, which is useful to estimate potential contributions to the radio emission from star formation or Active Galactic Nuclei (AGN) activity; 
\item Spectroscopic classifications that, together with the redshift information, can guide expectations for the detectability of possible radio counterparts for a given level of sensitivity. 
\end{itemize}

A downside of our VLA (GHz) follow-up strategy is that it may miss radio transients with faint optical counterparts, such as short GRB optical afterglows observed largely off-axis \citep{mb12}. Transients that are faint in the optical may still be identifiable via observations at lower radio frequencies ($100-150$\,MHz) by larger FOV radio facilities such as the Low-Frequency Array for radio astronomy \citep[LOFAR;][]{vwg13} and the Murchinson Widefield Array \citep[MWA;][]{bck13}, which indeed participated in the Advanced LIGO EM follow-up program during O1 \citep{abbot2016c}. We note, however, that GHz observations with the VLA offer some important advantages such as deep sensitivity, lower number of contaminating astrophysical sources, and less radio frequency interference (RFI).

For the three Advanced LIGO O1 GW alerts (GW150914, G194575, and GW151226; see Section \ref{intro}),  iPTF covered error areas of $\approx 126\,\rm\,deg^2$ \citep{kcs16}, $\approx 1114\,\rm\,deg^2$ \citep{gcn18497}, and $\approx 952\,\rm\,deg^2$ \citep{gcn18762}, respectively. In the case of GW150914, the prior probability that the iPTF imaged area contained the true location of the source was only 2.5\% due to  Sun angle and elevation constraints, and the iPTF observations started on the second night after the GW trigger \citep[UT Sep 17;][]{kcs16}. In the case of GW151226, the iPTF containment probability was 51\% for the total imaged area, and 37\% for the portion of the iPTF imaged area for which previously obtained reference images were available \citep{gcn18762}. iPTF observations of GW151226 started $\approx 1.9$\,days after the GW trigger \citep{gcn18762}.

In the areas imaged as part of the O1 triggers follow-up, iPTF discovered (on the same night the observations started) a total of 70 optical transients, out of which 16 were followed--up in the radio with the VLA (Table~\ref{tb:trigger_info}).  As we explain in what follows, no radio counterpart to the iPTF optical transients was found. However, for several of the iPTF candidates we discovered excess radio emission related to star formation in the host galaxy. 
Generally, we adhered to the criteria outlined in \citet{kcs16} for optical candidates rejection or spectroscopic / additional photometric follow--up.
For the VLA follow-up, we applied the following criteria:
\begin{itemize}
\item For optical candidates with same night optical spectral classification, we discarded SNe of type Ia,  SNe of type II, and AGNs, due to the low probability of these being strong GW emitters at cosmological distances. 
\item For optical candidates with same night optical spectral classification, we carried out at least one single-band  (typically C-band, centered around $\approx 6$\,GHz) continuum VLA observation of stripped-envelope core-collapse SNe (type Ib/c), and/or peculiar/rare SNe such as super-luminous and CSM-interacting ones (e.g., type Ibn).
\item When spectral classification was not immediately available, we carried out at least one single-band continuum VLA observation of optical transients coincident with galaxies at low redshifts, and/or unclassified optical candidates showing fast temporal evolution \citep[as possible kilonova candidates, see e.g.,][]{lp98,mb12,bk13,mf14}.
\item For optical candidates for which a radio excess was detected in the first radio epoch, for type Ic SNe  \citep[because of their connection to GRBs; e.g., ][]{wb06,cgk15}, for CSM-interacting SNe \citep[for which one might expect late-time radio emission;][]{ofc13,cog14,cgk15}, and for other optical transients of unclear nature, we carried our further VLA follow-up, via multi-epoch single-band (so as to constrain temporal variability) and/or multi-band (so as to constrain radio spectral properties) observations.
\item Finally, the statistic $\left|\Delta S/\sigma_{\Delta S} \right|\geq 4.3$ (with $\Delta S$  being the difference in flux between two epochs, and $\sigma_{\Delta S}$ the error on the flux difference) was used to establish whether radio flux variability was present in any of the candidates observed on at least two epochs.  The variability threshold was chosen so that the $\it t$--statistic lies beyond the 95\% confidence interval \citep[][]{mhb16}. Evidence for variability (although we found none in this search) is considered a reason for further multi-epoch multi-band VLA follow-up. 
\end{itemize}
 We discuss in detail the results of the VLA follow-up of iPTF transients in Sections \ref{DA} and \ref{radio_detected_iPTF}.

\begin{figure}
\begin{center}
\includegraphics[width=8.5cm]{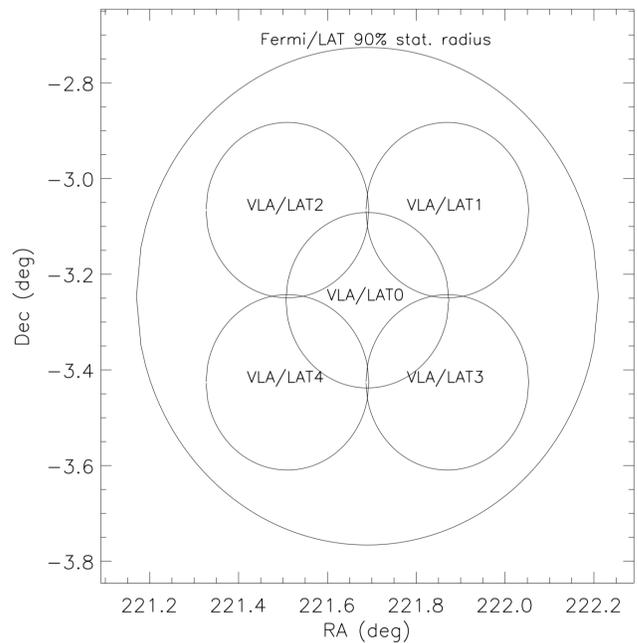} 
\vspace{-1.0cm}
\caption{Localization area of the possible \textit{Fermi}/LAT transient identified during the follow-up of G194575 (0.52\,deg 90\% statistical containment radius), and the VLA pointings used to cover it (VLA/LAT0-LAT4). Each of the VLA pointings are shown as circles of $\approx 11$\,arcmin radius, corresponding to $\approx 20\%$ of the primary beam at $3$\,GHz.}
\label{fig:pointings}
\end{center}
\end{figure}

For event G194575, in addition to selected iPTF candidates localized to arcsec positions, we also followed--up a possible $\gamma$-ray transient detected by the \textit{Fermi}/LAT in the GW localization area, and localized by it to $\approx 0.52$\,deg radius (plus $\approx 0.05$\,deg systematic error) at $\lesssim 16$\,deg from the Sun \citep{gcn18458}. Five VLA  pointings were carried out  at a central frequency of 3\,GHz (Fig.~\ref{fig:pointings}). These observations represented a first ``test run'' aimed at evaluating the feasibility of VLA follow--ups of limited regions of the GW localization area not accessible to optical observatories due to Sun constraints. The results of this run are reported in Sections \ref{DA} and \ref{Sec:Fermi}.

\section{Observations and data reduction}
\label{DA}
VLA follow-up observations of iPTF candidates in the GW error regions of the three O1 alerts were carried out at a central frequency of $\approx6$\,GHz, with a 2\,GHz nominal bandwidth, and with the array in its D or DnC configuration.
Multi--frequency observations were performed on selected targets (according to the criteria described in Section \ref{VLA_program}).
The Common Astronomy Software applications (\texttt{CASA}) was used to calibrate, flag, and image the data.
The automated VLA calibration pipeline for \texttt{CASA} was used to calibrate the raw data.
When needed, further flagging was carried out manually after visual inspection of calibrators and source data.

Follow-up observations of the possible \textit{Fermi}/LAT transient identified in the error region of G194575 (see Section \ref{eventG194575}) were carried out on two epochs at a central frequency of 3\,GHz, with the VLA in its D configuration. These data were not calibrated using the automated calibration pipeline due to RFI and elevated noise level caused by Sun proximity (the angular distance between the center of the \textit{Fermi}/LAT transient localization area and the Sun was $\approx 12.5$\,deg and $\approx 15.6$\,deg, respectively, on the two VLA epochs). After flagging, the usable data for the second epoch were limited to only two spectral windows, for a total nominal bandwidth of 256\,MHz. We thus restricted the first epoch analysis to these same two spectral windows.

\begin{figure}
\begin{center}
\includegraphics[width=6.8cm,angle=-90]{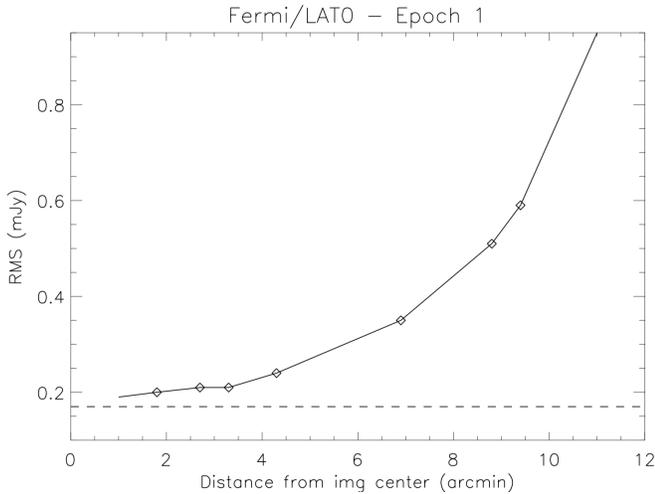}
\caption{RMS vs radius of the first epoch image of one of the five VLA fields covering the localization area of the possible \textit{Fermi}/LAT transient (LAT0; see Fig. \ref{fig:pointings}) . The dashed line marks the RMS of the residual image calculated without applying primary beam correction.   Diamonds mark the RMS values at the distances of the radio sources detected in this field (see Table \ref{tb:fermi}).  \label{fig:rms_vs_radius}}
\end{center}
\end{figure}

Images of the observed VLA fields were formed from the visibility data using the  \texttt{CLEAN} algorithm.
In creating the VLA images for the five fields covering the localization area of the possible \textit{Fermi}/LAT transient, during the \texttt{CLEAN} step, we corrected for the shape of the VLA primary beam  ($\approx 15$\,arcmin Full-Width-at-Half-Power). The primary beam correction was applied to a region extending up to $20\%$ power radius of the primary beam (which is the standard option in \texttt{CLEAN}). 
As expected, after applying the primary beam correction, the RMS of the residual image increases with increasing distance from the image center (Fig.~\ref{fig:rms_vs_radius}). 

In the VLA images collected as part of the follow-up of iPTF candidates, we searched for excess radio emission using a circular region centered on the candidates optical positions, with radius comparable to the nominal half-power synthesized beam width of the VLA for the frequency band at which the observations were carried out.  For the VLA images collected as part of the follow-up of the \textit{Fermi/LAT} transient, \texttt{BLOBCAT} \citep{hmc12} was used to construct a list of candidate sources (blobs) formed by locating pixels with surface brightness greater than $5\times$ the RMS noise (as estimated in the central region of the map), and then searching for neighboring pixels having surface brightness greater than $3\times$ the RMS. The blobs were visually inspected, and their significance was corrected for the shape of the primary beam by rescaling the estimated RMS according to source distance from the image center (see Fig. 1).  Candidate sources with peak-flux-to-primary-beam-corrected RMS ratio less than 5 were discarded.

In Tables~\ref{tb:event1}-\ref{tb:fermi} we list the radio detections (and/or 3$\sigma$ upper-limits for the iPTF follow-up) obtained during our O1 campaign.  In these Tables, upper-limits for the non-detections are calculated as $3\times$ the noise RMS around the map center, close to the location of the iPTF transient. For the detections, we report peak fluxes estimated using the  \texttt{IMSTAT} routine. Flux errors are calculated as the quadratic sums of the RMS map error, and a 5\% fractional error that accounts for absolute flux calibration errors \citep{wsp86,cog14}. The VLA positions were determined using the \texttt{imfit} routine in \texttt{CASA}, which fits an elliptical Gaussian component on an image region which was chosen as a circular region of radius equal to the half-power width of the major axis of the actual VLA synthesized beam. Position errors were estimated using both \texttt{imfit} (RA and Dec errors returned for the chosen extraction region were added in quadrature), and the half-power width of the semi-major axis of the VLA synthesized beam, divided by the peak-flux-to-RMS ratio. These position error estimates were usually found to be in good agreement for point sources with fluxes above $\approx \times 5$ the map RMS at the source position. Positions errors smaller than 0.1\,arcsec were set to this value to account for the VLA systematic positional uncertainty. In Tables~\ref{tb:event1}-\ref{tb:fermi}, we conservatively report the largest of these error estimates as our position error radii.

In Tables~\ref{tb:event1}-\ref{tb:event3}  we also report the iPTF classification (column labelled ``Class") of the various transients. Transient classifications followed by a question mark are less secure than unmarked ones. All of the iPTF transients with a nearby radio detection were also observed/classified spectroscopically (see \citet{k16} and \citet{scs16} for further details).

\begin{table*}
\begin{center}\begin{footnotesize}
\caption{VLA follow-up observations of iPTF transients identified in the error area of event GW150914. Data taken before (after) MJD 57386 are with the VLA in its D (DnC) configuration.   $\Delta T$ is the time between the iPTF discovery and the VLA observation.\label{tb:event1}}
\begin{tabular}{cccccccc}
\hline
\hline
name &  RA  Dec (iPTF) &  Class& VLA epoch & $\Delta T$ & Freq. & Flux or 3$\sigma$ UL  \\ 
& (hh:mm:ss~~deg:mm:ss) & & (MJD) & (day) & (GHz)& ($\mu$Jy beam$^{-1}$)& \\
\hline
iPTF15cyk&07:42:14.87~~20:36:43.4 &SLSN I&     57310.5  &     28 &  5.4&  $\lesssim$30  &\\
"&"&"&      57362.2 &   80 &    5.4&  $\lesssim$23  &\\
"&"&"&      57407.1&    125 &  5.4&  $\lesssim$23  &\\
\hline
\end{tabular}
\end{footnotesize}\end{center}\end{table*}

\begin{table*}
\begin{center}\begin{footnotesize}
\caption{VLA follow-up observations of iPTF transients identified in the error area of trigger G194575. All observations were carried out with the VLA in its D configuration. $\Delta T$ is the time between the iPTF discovery and the VLA observation. Radio sources offsets with respect to the iPTF positions are reported in the second to last column. } 
\label{tb:event2}
\resizebox{18cm}{!}{
\begin{tabular}{lllcccllll}
\hline
\hline
name &  RA  Dec (iPTF)   & Class & VLA epoch & $\Delta T$ & Freq. & Flux or 3$\sigma$ UL & RA  Dec (VLA) &offset & pos. err. (VLA) \\ 
& (hh:mm:ss~~deg:mm:ss)&  & (MJD) & (day) & (GHz)& ($\mu$Jy beam$^{-1}$)& (hh:mm:ss~~deg:mm:ss)&(arcsec)&(arcsec) \\
\hline
iPTF15dkk&23:50:17.21~~-03.09.59.8&   Nuclear?     &  57324.1    &   6 &   6.3&$\lesssim60$ & - & - & -\\
"                &                                        &     -      &   57330.1   &     12 &  6.4  &    $\lesssim45$ & - & - & -\\
\hline
iPTF15dkm& 23:37:17.70~~-03.31.57.1 &SN II&      57330.1&    12 &  6.4& $\lesssim 51$ & - & - & - \\
\hline
iPTF15dkn&23:50:11.10~~+00:05:47.8&    Nuclear?  &   57324.1&   6 &   6.3&   $\lesssim66$ & - & - & -\\
"                &   "                                      &    " &   57330.1&     12&  6.4    &   $\lesssim69$ & - & - & -\\
\hline
iPTF15dkv& 00:31:05.08~~-02:39:08.2& RadioS &      57324.1&      6 & 5.2&       $(4.91\pm0.25)\times10^3$ &00:31:04.952~~-02:39:12.62  &   4.8   & 0.1\\
"                &         "                               & - &    "                 &      " & 7.4&       $(3.42\pm0.17)\times10^3$ &00:31:04.951~~-02:39:12.58 &      4.8&  "\\
"                &          "                                & - &    57330.1&      12 & 5.2&       $(4.79\pm0.24)\times10^3$ & 00:31:04.979~~-02:39:12.60     & 4.6&   " \\
"                &          "                                & - &     "               &      " & 7.4&       $(3.26\pm0.16)\times10^3$ &  00:31:04.975~~-02:39:12.66&  4.7&    "\\
\hline
iPTF15dld&00:58:13.28~~-03:39:50.3  &BL-Ic&     57324.1 &      6 & 6.4&       $69\pm21$    &00:58:13.094~~-03:39:49.18&   3.0   &     2.1\\
"               &                   "                        &   "     &    57330.1 &     12 & "    &       $90 \pm18$   &00:58:13.155~~-03:39:48.63&  2.5 &     1.1\\
"               &                   "                        &   "     &     57336.0      &  18   &      6.4 &       $86\pm13$   & 00:58:13.035~~-03:39:47.58     &  4.6    &     1.5\\
"               &                   "                        &   "     &     57336.1  &  18 & 2.9&       $\lesssim 264$ &-&     -&      -\\
"               &                   "                        &   "     &     "            &      " & 6.4&       $81\pm17$     &00:58:13.110~~-03:39:47.68&      3.6&      1.8\\
"               &                   "                        &   "     &     "            &      " & 9.0&       $63\pm14$ &00:58:13.234~~-03:39:48.92&      1.5&      1.1\\
\hline
iPTF15dlj&01:19:02.87~~+10:00:04.8&SN II&     57324.1&  6 &    6.3&$\lesssim60$ & - & - & - \\
"              &                "                    & "     &     57330.1&     12 &  6.4   &$\lesssim57$ & - & - &- \\
\hline
iPTF15dln&00:58:19.67~~+07:14:05.0&Nuclear?&       57324.1&  6 &    6.4&$\lesssim 90$ &- & - &- \\
"                &                   "                    & "          &      57330.1&    12 &   "   &$\lesssim 72$ & - & - &- \\
\hline
iPTF15dmk&01:24:54.37~~+00:37:07.5&SN II&      57324.1&  6 &     6.4&$\lesssim  132$ & - & - & - \\
"                  &                    "                  & "     &    57330.1&    12 &   " &$\lesssim 132$ &  - & - & - \\
"                  &       "                               &  "     &     57336.1    & 18 & "     &$\lesssim96$ & - & - & - \\
\hline
iPTF15dmn& 00:28:56.73~~-11:24:19.8 & AGN&      57324.1&     6 &  5.1&       $(9.43\pm0.61)\times10^2$       &00:28:56.786~~-11.24.19.72&  0.83   &    0.39\\
"                  &                                            &     "  &      "               &     6 &  7.4&       $(6.03\pm0.43)\times10^2$       &00:28:56.774~~-11:24:20.22&     0.76&    0.37\\
\hline
iPTF15dmq&   23:35:30.74~~+06:27:01.1 & Asteroid & 57324.1     &   6 &   6.3&      $\lesssim78$  & - & - & - \\
"              &   "                                          & "             & 57330.1    &    12 &   6.4&     $\lesssim72$  & - & - & - \\
\hline
iPTF15dmu& 01:30:23.01~~-04:24:36.7 & SN?&       57336.1&    18 & 5.2&       $(4.61\pm0.32)\times10^2$ & 01:30:23.128~~-04.24.36.95&     1.80 &     0.51\\
"                  &                      "                    & " &               "         &      " & 7.5&       $(3.19\pm0.24)\times10^2$&     01:30:23.112~~-04:24:37.09 &  1.58 & 0.44   \\
"                  &                      "                    & " &      57344.0 &  26 & 2.7&   $(8.74\pm0.49)\times10^2$& 01:30:23.293~~-04.24.39.64&     5.15\footnote{Two unresolved emission components present.} &     0.76\\
"                  &                      "                    & " &                   "      &    26 &  5.2&      $(5.08\pm0.29)\times10^2$&01:30:23.220~~-04:24.38.31&      3.62\footnote{Two marginally resolved emission components present.}&     0.67\\
"                  &                      "                    & " &                   "      &   26 &    7.4&      $(3.17\pm0.23)\times10^2$&01:30:23.095~~-04:24:36.62&      1.26&     0.43\\
"                  &                     "                     & " &                   "      &   26 &   8.5&      $(2.62\pm0.21)\times10^2$&01:30:23.052~~-04:24:36.29&     0.76&     0.54 \\
"                  &                     "                     & " &                   "      &   26 &   9.5&      $(2.44\pm0.19)\times10^2$&01:30:23.032~~-04:24:36.22&     0.58&     0.33\\
\hline
iPTF15dnh&00:59:38.27~~-14:11:56.8 & -  &   57324.1&   6 &   6.3&$\lesssim123$ & - & - & - \\
"                &       "                                  & -  &     57330.1&   12&    6.4    &$\lesssim159$ & - & - & - \\ 
\hline
iPTF15dni&01:25:04.02~~-04:42:30.4 &AGN? &     57324.1 &   6 &   5.1&       $(1.479\pm0.081)\times10^3$ &01:25:04.019~~-04:42:30.87&      0.47&  0.34\\
"               &                     "                    &  "        &      "               & "          &      7.4&       $(1.018\pm0.064)\times10^3 $    &01:25:03.998~~-04:42:31.28&      0.94&    0.36\\
"               &                     "                    &  "        &     57330.1  &    12 &  5.2&       $(1.465\pm0.080)\times10^3$   &01:25:04.031~~-04:42:31.34&      0.97&   0.30\\
"               &                     "                    &  "        &        "             & "          &      7.4&       $(1.002\pm0.059)\times10^3$   &01:25:04.007~~-04:42:31.37     &     1.01 &      0.26\\
\hline
\end{tabular}}
\end{footnotesize}\end{center}\end{table*}

\begin{table*}
\begin{center}
\begin{footnotesize}
\caption{VLA follow-up observations of iPTF transients identified in the error area of event GW151226. All the observations were carried out with the VLA in its DnC configuration. $\Delta T$ is the time between the iPTF discovery and the VLA observation. Radio sources offsets with respect to the iPTF positions are reported in the second to last column.  \label{tb:event3}}
\begin{tabular}{lclccclcll}
\hline
\hline
name & RA  Dec (iPTF)&  Class & VLA epoch & $\Delta T$ & Freq. & Flux or 3$\sigma$ UL & RA  Dec (VLA)& offset & pos.err. (VLA)\\ 
& (hh:mm:ss~~deg:mm:ss)& & (MJD) & (day) & (GHz)& ($\mu$Jy beam$^{-1}$)& (hh:mm:ss~~deg:mm:ss) &(arcsec)&(arcsec) \\
\hline
iPTF15fgl&02:32:59.78~~+18:38:07.7&SN Ibn&       57395.2&    11&  6.3&     $58.3\pm9.0$ &02:32:59.850~~+18.38.07.01&      1.22& 0.90\\
"	      &         "                             & "     &       57400.1&      16 & 3.1&      $86\pm27$ & 02:32:59.870~~+18:38:07.56 &   0.72   & 2.9 \\
"	      &         "                             & "     &             "          &    " &  9.0&      $36.3\pm7.5$     &      02:32:59.830~~+18:38:07.35 &   0.79   & 0.75     \\
"	      &         "                             & "     &            "          &     " & 14.8&     $25.2\pm6.6$      &    02:32:59.907~~+18:38:07.10   &    1.9  &  1.0    \\ 
"	      &         "                             & "     &       57401.1&      17 & 6.3&       $52.8\pm9.1$ & 02:32:59.855~~+18:38:06.91&      1.33&     0.91\\
"	      &         "                             & "     &       57407.1&     23 & 6.4&       $54.8\pm8.5$ &    02:32:59.901~~+18:38:07.12&     1.80 &   0.81  \\
"	      &         "                             & "     &     57409.1 &    25 &  3.0&       $97\pm28$ &02:32:59.800~~+18:38:04.76&      3.0&      4.6\\
"	      &         "                             & "     &          "                & "   &      9.0&       $31.2\pm7.2$ &02:32:59.879~~+18:38:07.55&      1.4&      0.68\\
"	      &         "                             & "     &     "                      & "  &      14.7&       $28.3\pm6.7$ & 02:32:59.945~~+18:38:08.50&      2.5&    1.2 \\
\hline
iPTF15fhl&12:28:13.60~~17:37:01.4& SN Ib/c&       57386.4&     2 & 5.1&       $365\pm22$ &12:28:13.704~~+17:36:53.57&      7.96&     0.34\\
"             &                      "                  & " &       "                 &      " & 7.5&       $231\pm20$ &12:28:13.664~~+17:36:53.43&      8.03&     0.42\\
"             &                     "                  & " &       57394.4&      10 & 5.1&       $328\pm24$ &12:28:13.711~~+17:36:54.33&      7.24&     0.39\\
"             &                     "                  & " &       "                &      " & 7.4&       $206\pm17$ & 12:28:13.733~~+17:36:53.87&      7.78&     0.35\\
"             &                     "                  &" &        57408.3 &      24 & 5.2&       $414\pm31$ & 12:28:13.783~~+17:36:54.32&      7.56&      0.94\\
"             &                     "                  &" &       "                  &      " & 7.4&       $322\pm27$ &12:28:13.720~~+17:36:54.18&      7.42&      0.62\\
\hline
\end{tabular}
\end{footnotesize}\end{center}\end{table*}

\subsection{GW150914}
GW150914 was a binary BH coalescence \citep{abbot2016b}. BH component masses and redshift are reported in Table \ref{tb:trigger_info}.  A super-luminous SN (iPTF15cyk) was the most notable optical transient discovered in the error region of GW150914 by the iPTF. This SN, although unrelated to the GW trigger, was followed-up in radio with the VLA on three epochs covering a timeframe of 1-3\,months after the iPTF discovery (as listed in Table~\ref{tb:event1}). The timescale of our VLA monitoring was motivated by models predicting possible late-time radio emission from super-luminous SNe \citep{ofc13}. No radio emission was detected at the location of iPTF15cyk. We refer the reader to \citet{kcs16} for a detailed description of the iPTF follow-up performed for GW\,150914.

\subsection{G194575}
\label{eventG194575}
G194575 was the second GW alert sent out during Advanced LIGO O1 \citep{gcn18442}, subsequently declared as an event of no interest. Our VLA follow-up of optical transients identified in the error region of G194575 by the iPTF covered a timeframe of approximately 1 week to 25\,days after the iPTF discovery (Table~\ref{tb:event2}), optimized for the detection of radio emission from off-axis short GRBs and relativistic SNe, and covering the rising part of the emission potentially associated with slower ejecta from NS-NS (or NS-BH) binaries \citep[see Section \ref{conclusion} and][]{cgk15}. 

Two iPTF sources with nearby radio detections in the error areas of G194575 are most notable, iPTF15dld and iPTF15dmu. iPTF15dld was a broad-lined Type Ic (BL-Ic) SN at a redshift of $z=0.047$, especially interesting because of the rarity of these type of naked core-collapse SNe \citep[see e.g.][and references therein]{cgk15}. Given the connection between BL-Ic SN and GRBs, this object was also followed-up in X-rays with \textit{Swift/XRT} at 0.3--10 keV, under our BL-Ic program\footnote{\textit{Swift} Cycle 11 proposal $\#1114155$; PI: Corsi}. The \textit{Swift/XRT} observation resulted in a non--detection \citep{cgk15}. For iPTF15dmu, iPTF provided  photometric observations showing lack of significant fading on a two-week timescale, the disappearance of the transient on a two-month timescale, and an off-nuclear location. These properties are suggestive of a SN origin. 

Out of the total 13 iPTF transients followed--up with the VLA for G194575, 5 showed a significant ($\gtrsim 3\sigma$) radio excess near the iPTF location (Table~\ref{tb:event2}), and were re--observed on multiple epochs to search for variability.  According to the variability statistic defined in Section \ref{VLA_program}, none of the iPTF transients with a nearby radio detection (and more than one radio observation) showed any variability over the timescales of our follow-up. The detected radio excesses are easily accounted for by star formation of the host galaxies, and none of them are attributed to radio counterparts of the iPTF transients  (see Section~\ref{radio_detected_iPTF}). 

Within the error area of G194575, $\sim 10^3$\,s after the GW trigger, a possible $\gamma$-ray transient was detected by the \textit{Fermi}/LAT \citep{rbg16}. As described in Section \ref{VLA_program}, we observed five fields partially covering the \textit{Fermi}/LAT localization area for this $\gamma$-ray trigger (Fig. \ref{fig:pointings}). Each of the fields was observed on two epochs separated by about $\approx 8$\,d. Sources were extracted from the first epoch images as described in Section \ref{DA}. 
Their locations were recorded and used to measure the fluxes in the second epoch images (see Table~\ref{tb:fermi}). As we discuss in Section 5, almost all of these sources were previously cataloged radio sources and none were found to be variable over the two epochs of our follow-up. 

\subsection{GW151226}
GW151226 is the second significant binary BH merger detected by Advanced LIGO during O1. Component  BH masses and redshift range for this event are reported in Table \ref{tb:trigger_info}. Similarly to the case of G194575, our VLA follow-up of optical transients identified in the error region of GW151226 by the iPTF covered a timeframe of approximately 1 week to 25\,days after the iPTF discovery (Table~\ref{tb:event3}). Two iPTF transients found in the error region of GW151226 were most notable (although likely unrelated to the GW event given its binary BH merger origin), iPTF15fgl \citep[also named PS15dpn, a type Ibn SN; see ][for more details]{scs16}, and iPTF15fhl (a Type Ib/c SN). These transients were both followed up with the VLA. While significant radio excesses were detected close to the locations of iPTF15fhl and iPTF15fgl (Table~\ref{tb:event3}), these radio detections are easily accounted for by star formation in the host galaxies of these SNe. Indeed, according to the variability statistic defined in Section \ref{VLA_program},  none of the detected radio excesses showed any variability over the timescales of our follow-up (see Section~\ref{radio_detected_iPTF}). We thus find no evidence for radio counterparts to the SNe iPTF15fgl and iPTF15fhl.
\section{\lowercase{i}PTF candidates with radio detections}
Star-forming galaxies and  AGN represent the two major populations of radio sources expected to be found in extragalactic radio (cm) continuum surveys \citep[e.g.,][]{Condon1992,smj99, Smolcic2008, Smolcic2016}, and as such they can be a source of false positives in radio follow-up of GWs. Recent results indicate that below flux densities of $\sim 200\,\mu$Jy at 3\,GHz, star-forming galaxies begin to dominate in terms of fractional contribution to the total source sample \citep{Smolcic2016}, although low luminosity AGN may also be present \citep{mfo13}.  At the mJy level, the transient/variable radio sky is dominated by AGN \citep[]{smj99} with accretion onto a central BH, or supermassive binary BHs. Optical properties, radio spectral index, luminosity, and mid-IR properties are some of the tools that can be used to distinguish between e.g. variable AGN radio emission and catastrophic stellar explosions \citep[see also][]{mfo13}. 

\begin{figure*}
\begin{center}
\hbox{
\hspace{0.1cm}
\includegraphics[width=6.cm]{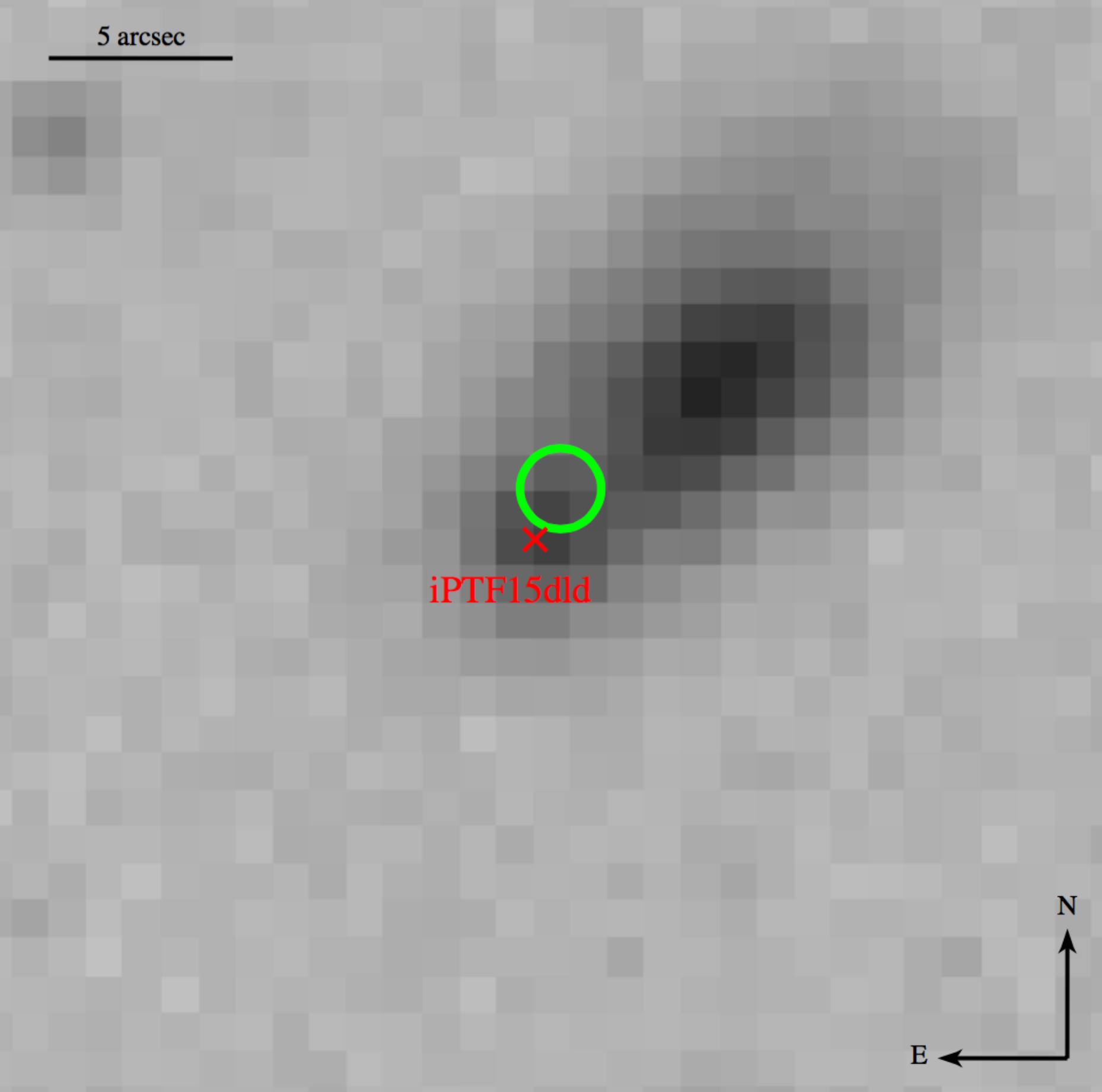}
\includegraphics[width=6.cm]{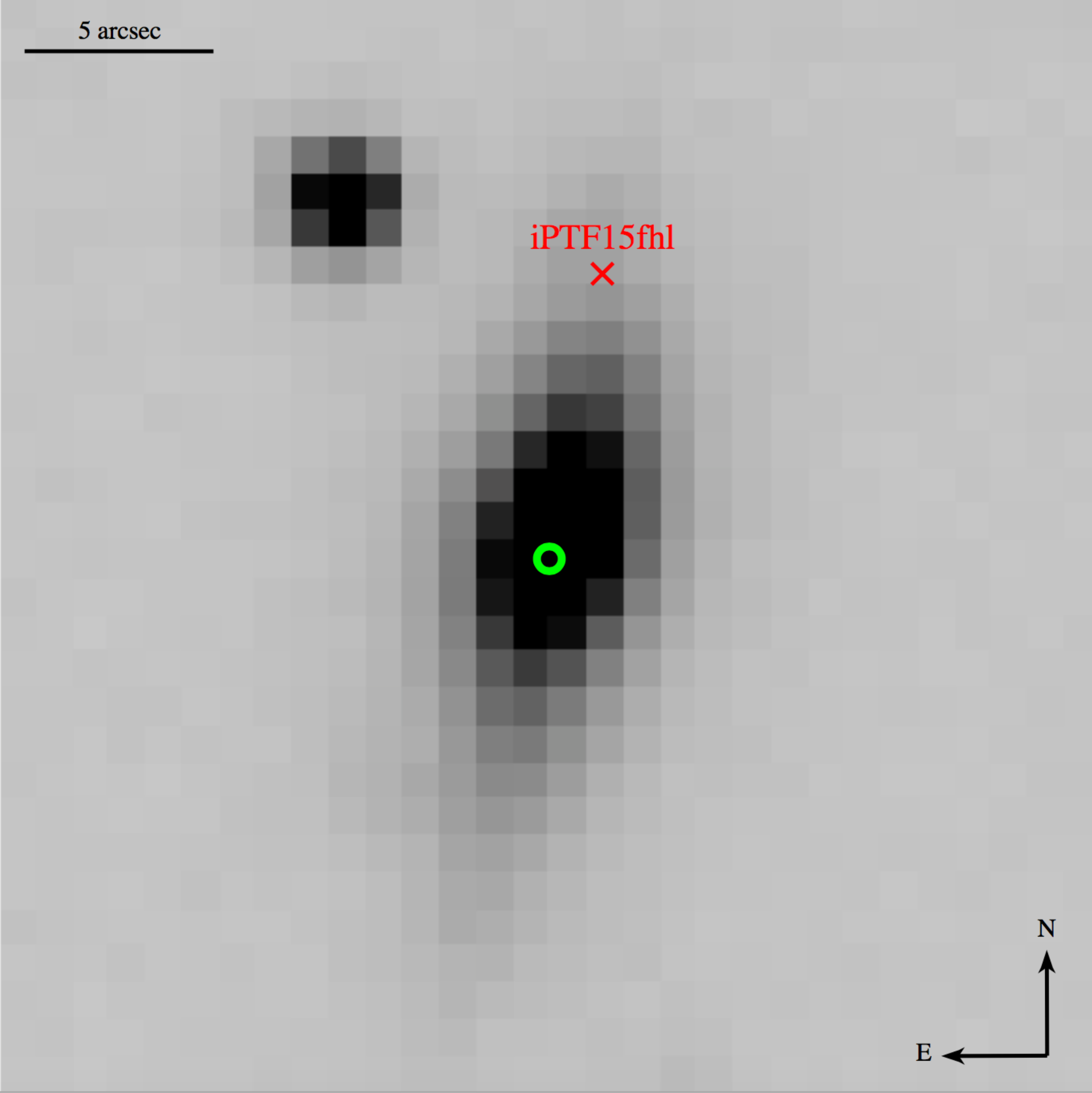}
\includegraphics[width=6.cm]{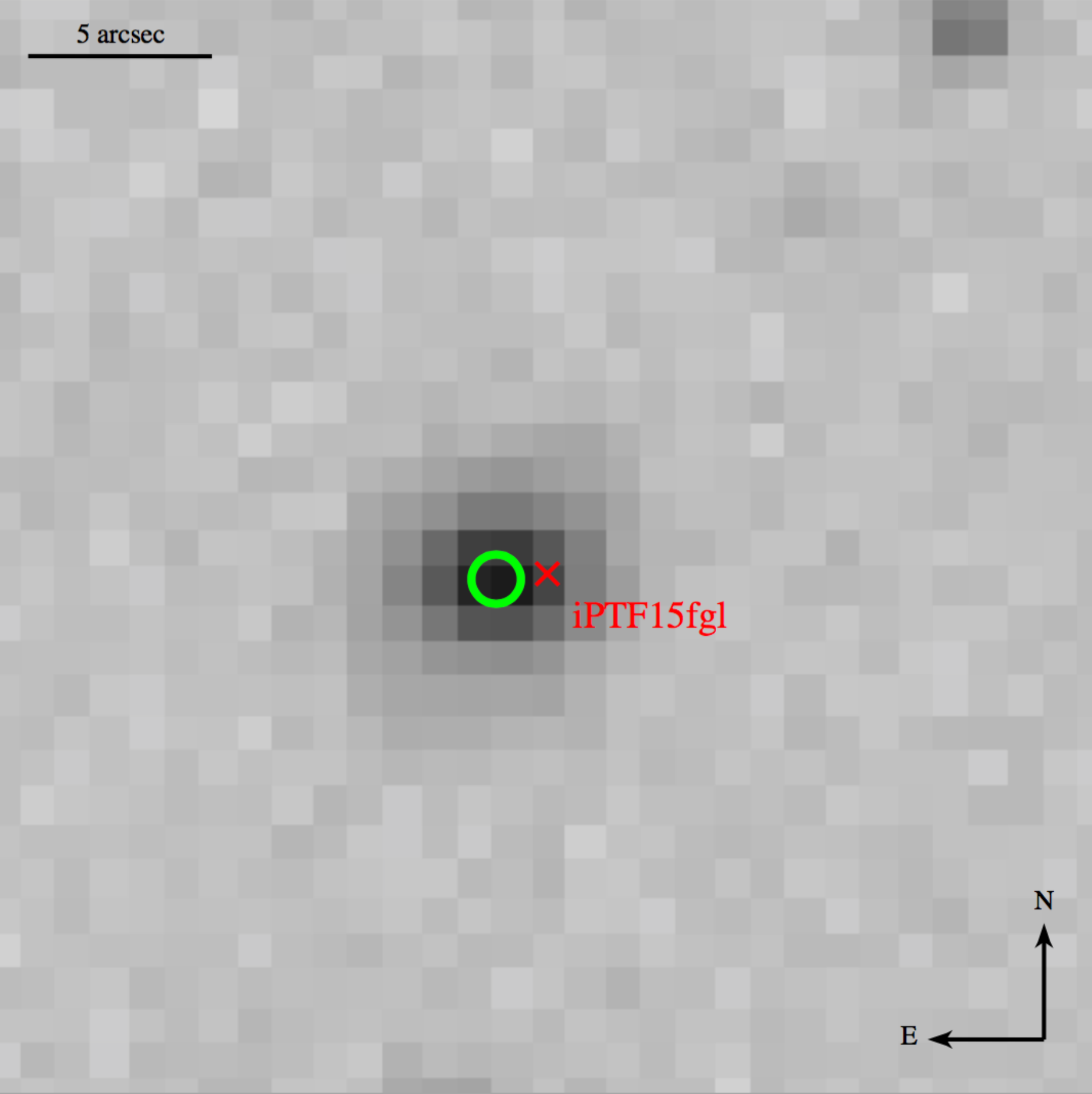}
}
\vspace{0.5cm}
\hspace{0.1cm}
\hbox{
\includegraphics[width=6.05cm]{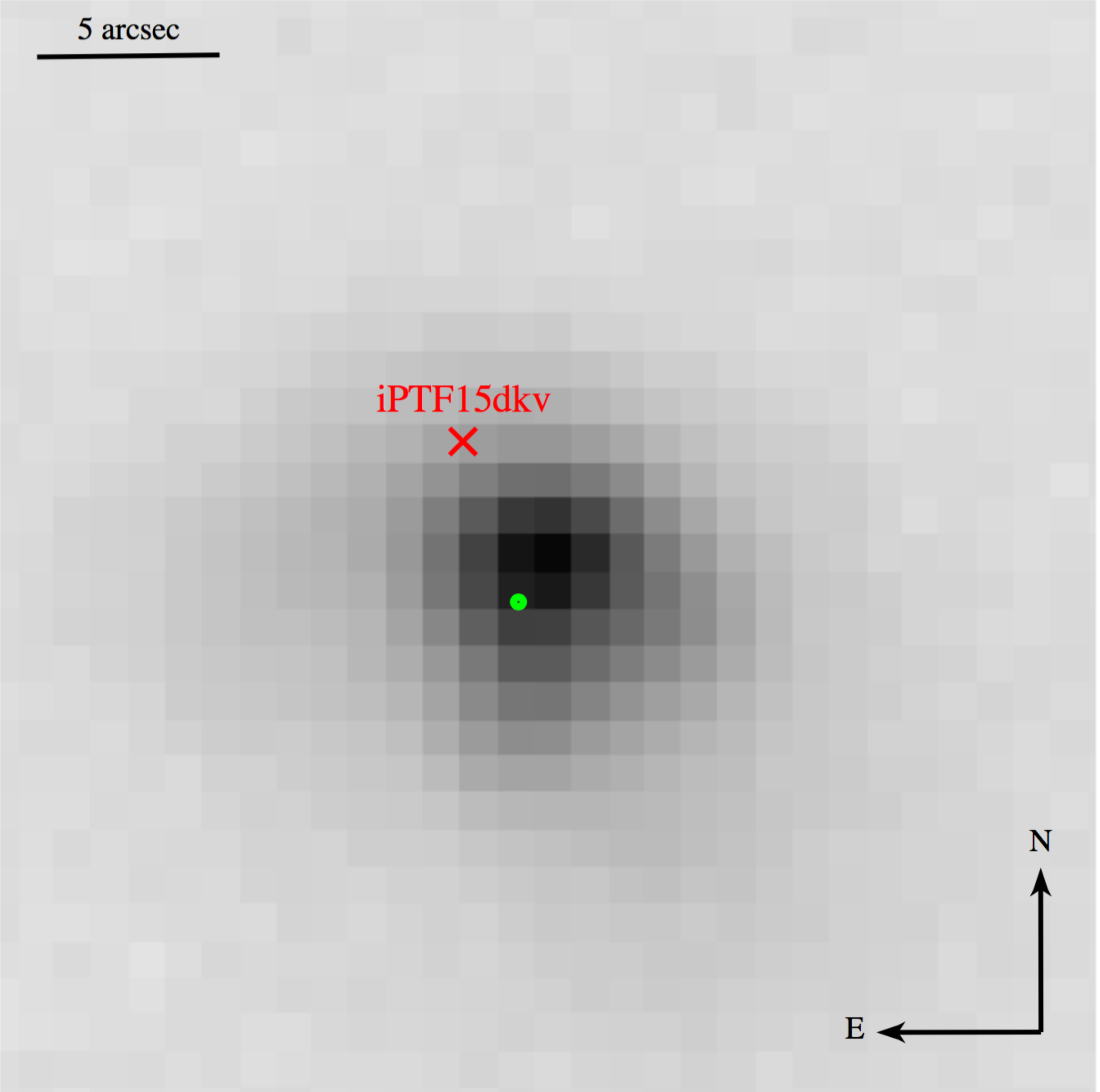}
\hspace{0.5cm}
\includegraphics[width=6.02cm]{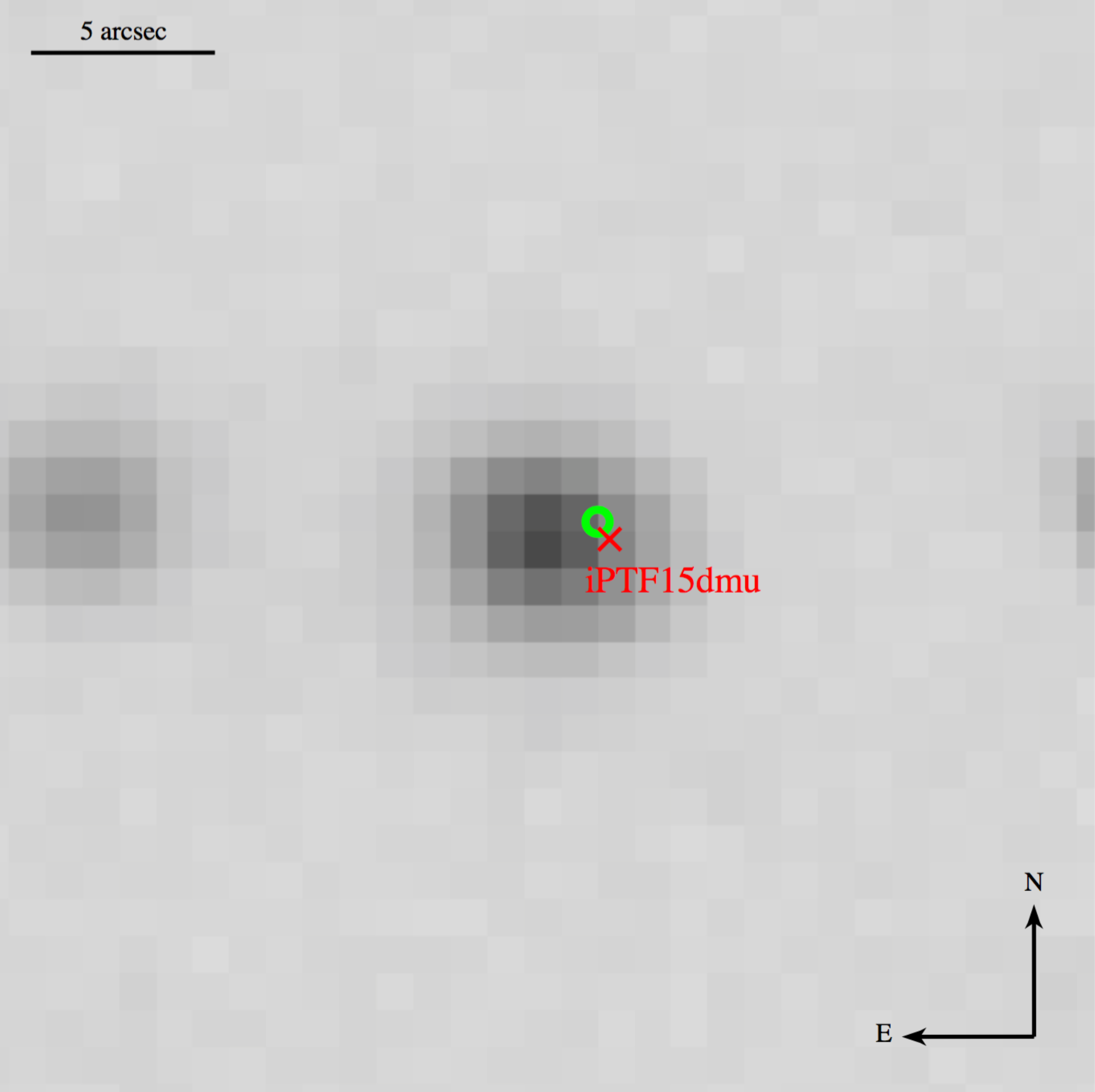}}
\vspace{0.5cm}
\hspace{1.cm}
\hbox{
\includegraphics[width=6.0cm]{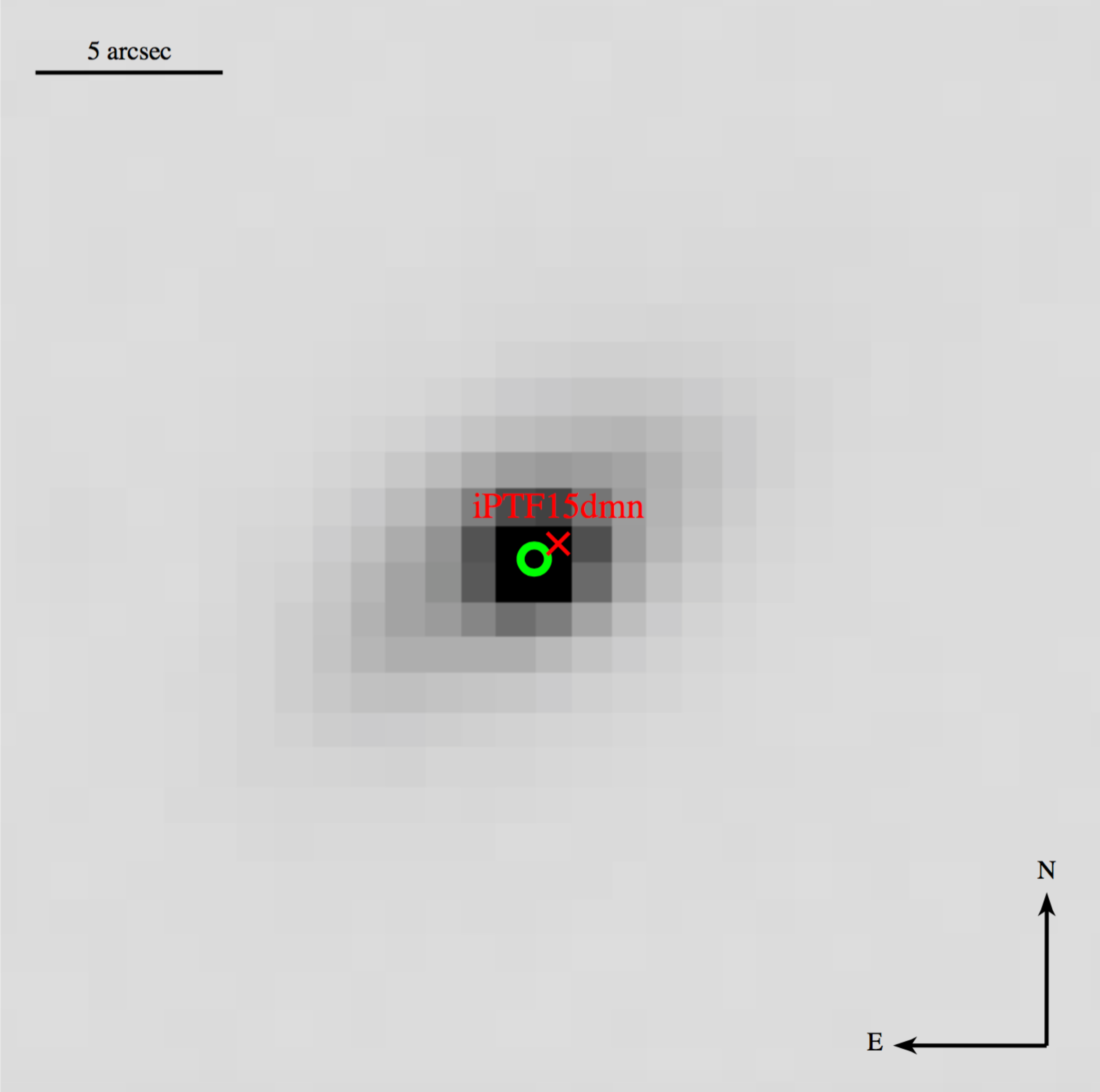}
\hspace{0.5cm}
\includegraphics[width=6.0cm]{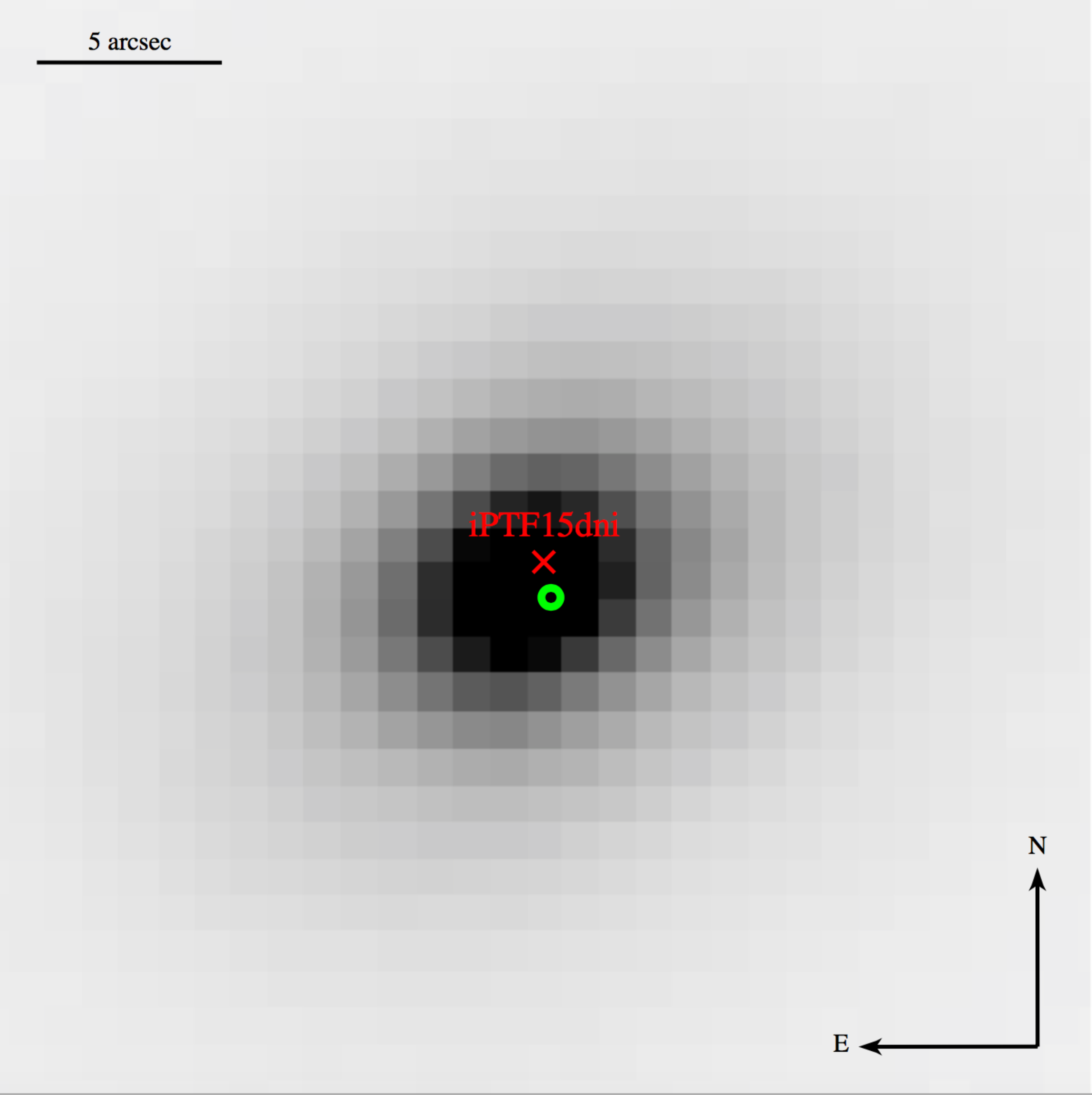}}
\caption{iPTF P48 R-band reference images of the host galaxies of: the BL-Ic SN iPTF15dld (top-left; follow-up of G194575); the Type Ic SN iPTF15fhl (top-center; follow-up of GW151226); the Type Ibn iPTF15fgl (top-right; also PS15dpn, follow-up of GW151226); the transients iPTF15dkv (center-left; follow-up of G194575) and iPTF15dmu (center-right; follow-up of G194575); and the AGN-associated transients iPTF15dmn (bottom-left; follow-up of G194575) and iPTF15dni (bottom-right; follow-up of G194575). The locations of the iPTF optical transients are marked with a red $\times$. The green circles show the locations (and estimated positions errors) of the detected VLA excesses (for each source, we center the green circle on the radio position derived from the observation affected by the smallest radio position error). In all cases, the radio detections are consistent with emission related to star formation in the host of the iPTF transients, and no variability in the measured radio fluxes is found over the timescales of our follow-up. We thus find no evidence for any radio counterpart to the optical transients themselves. See text for discussion. \label{figref1}}
\end{center}
\end{figure*}

For 7 of the 16 iPTF optical transients followed-up with the VLA during Advanced LIGO O1, we detected a significant radio excess in the vicinity of the iPTF transient location. As previously discussed, none of these radio detections showed evidence for radio variability over the timescales of our observations (see Section \ref{DA}). Because all of the iPTF transients with a nearby radio detection are embedded in host galaxies located within $z\lesssim 0.2$ (Fig. \ref{figref1}), and in the light of the above considerations, in this Section we study their radio-mid-IR-optical properties to gain insight into the origin of the detected radio excess. As we explain in what follows, it is reasonable to attribute the detected radio excesses to star formation (SF) from normal galaxies. However, weak AGN emission cannot be securely excluded given that AGNs span a large range in radio power. Moreover, for two of the hosts, an AGN contribution is evident in the optical band. 
\label{radio_detected_iPTF}
\subsection{Radio emission properties}
Most of the cm radio emission from normal galaxies is due to synchrotron radiation from relativistic electrons accelerated by SN explosions; a smaller contribution arises from free-free scattering. Star formation rates (SFRs) of normal galaxies may be estimated as \citep{mur11,pp13}:
\begin{equation}
\left(\frac{\rm SFR_{radio}}{\rm M_\odot\,yr^{-1}}\right) = 6.35\times10^{-29}\left(\frac{L_{1.4\,\rm GHz}}{\rm erg\,s^{-1}Hz^{-1}}\right),\label{eq:1}
\end{equation}
where $L_{1.4\,\rm GHz}$ is the luminosity at 1.4 GHz. We stress that the above Equation assumes that all of the detected radio emission is due to SF.

Table~\ref{tb:_SFR} lists the redshift, distance, estimated spectral index ($f_{\nu}\propto \nu^{\alpha}$),  and 1.4 GHz power, as well as estimated host galaxy SFR, for the iPTF transients with a detected radio excess close to the iPTF location. Radio spectral indices are estimated using fluxes measured in the two adjacent VLA sub-bands (for sources that only have single-band observations) averaged over the multiple epochs of observation, or by fitting the multi-band data (for sources that do have multi-band observations; see below). SFRs are  calculated using Eq. (\ref{eq:1}) and the 1.4\,GHz fluxes extrapolated from our VLA observations using the derived values of the spectral indices. 

\begin{figure}
\begin{center}
\includegraphics[width=8.8cm]{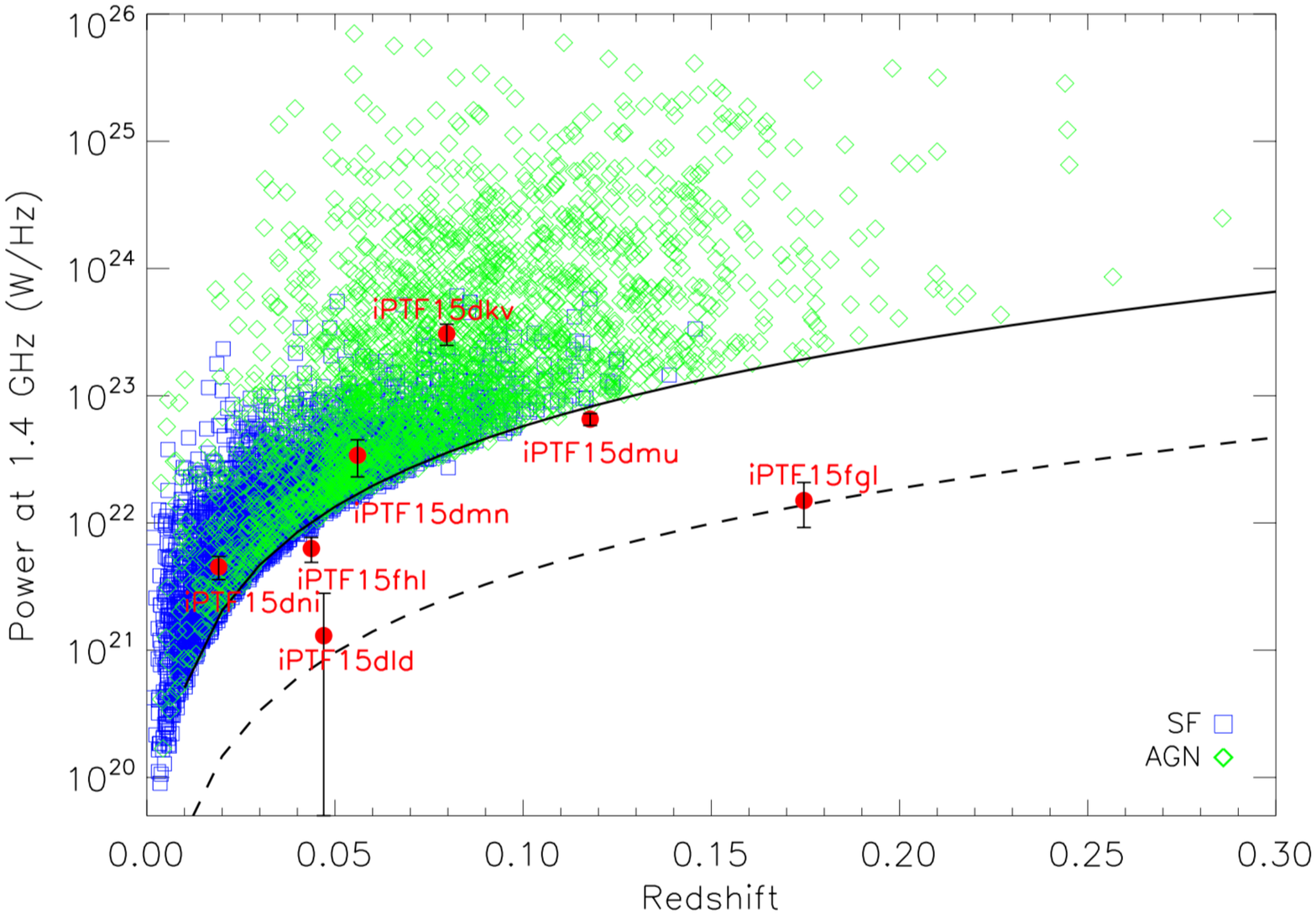}
\caption{Radio power at 1.4\,GHz versus redshift of AGN (diamonds) and SF galaxies (squares) listed in \citet{Mauch2007}. 
The extrapolated 1.4\,GHz fluxes from the host galaxies of the 7 iPTF transients with nearby radio detections (see Table \ref{tb:_SFR}) are marked as red filled circles. 
The solid curve marks the approximate $5\sigma$ sensitivity ($\approx 5\times0.45$\,mJy) of the NVSS survey \citep{ccg98}, and the dashed curve marks the extrapolation to 1.4\,GHz of the approximate $3\sigma$ sensitivity of our search at 5\,GHz  ($\sim 3\times(1.4\,\rm{GHz}/5\,\rm{GHz})^{-1}\times15\,\mu$Jy). \label{power_vs_z}}
\end{center}
\end{figure}

The SFRs derived from our radio detections are consistent with radio emission from normal galaxies \citep{Condon1992}. This is also shown in Figure~\ref{power_vs_z}, where we compare the extrapolated 1.4 GHz radio power as a function of redshift with the AGN and star-forming galaxies listed in \citet{Mauch2007}. (We note that our search extended to the  $\sim 10 \mu$Jy level, so we were sensitive to fainter objects than the ones included in \citet{Mauch2007}). Indeed, almost all of the host galaxies of the iPTF transients followed with the VLA are within the region of normal galaxies \citep{mac12,ghj14}. The excess radio emission detected in the host of iPTF15dkv, which matches a previously cataloged radio source, is the brightest of our sample and could be compatible with AGN emission. However, combining radio data with optical ones from the 6 degree Field Galaxy Survey \citep{jsc+04}, \citet{Mauch2007} have classified the host of iPTF15dkv as a SF galaxy rather than an AGN. In fact, \citet{Mauch2007} found that although the median radio power of AGN in their sample (log$(L_{1.4\,\rm{GHz}}$\,[W\,Hz$^{-1}$])=23.04) is almost an order of magnitude higher than the median for SF galaxies, AGN span a wide range in radio power ($10^{21}-10^{26}$\,W\,Hz$^{-1}$) and so it is hard to separate SF galaxies from AGN based \textit{solely} on a radio power cut.

\begin{table*}
\begin{center}\begin{footnotesize}
\caption{Transient name, classification, spectroscopic redshift, distance, derived radio spectral index $\alpha$, extrapolated 1.4 GHz flux, measured 1.4 GHz flux (from VLA First or NVSS), extrapolated host galaxy radio SFR, and host galaxy optical  SFR for the iPTF candidates with nearby VLA detections. See text for discussion.  \label{tb:_SFR}
}
\begin{tabular}{lllcccccl}
\hline
\hline
name  &Class & z&Distance& $\alpha_{\rm radio}$ & $L_{1.4\rm GHz}$ & VLA First 1.4 GHz & Radio SFR  & Optical SFR\\
          & & &(Mpc)  &                                       & ($10^{22}$\,W\,Hz$^{-1}$) &  ($10^{22}$\,W\,Hz$^{-1}$)& (M$_\odot \rm yr^{-1}$) & (M$_\odot \rm yr^{-1}$) \\
\hline
iPTF15dkv&      RadioS & 0.0797&      364&    $-1.06\pm0.14$     &    $30.7\pm5.8$&  $24.0\pm1.5\footnote{Includes 5\% absolute flux calibration error.}$&  $195\pm37$ & $\sim 51$\\
iPTF15dld&     SN BL-Ic & 0.047&      210&     $-0.76\pm0.72$ &        $0.13\pm0.15$ &-- & $ 0.86\pm0.95$& $\sim 0.31$ \\
iPTF15dmn&       AGN & 0.056&      252&    $-1.20\pm0.26$&       $3.4\pm1.1$ & -- &$21.4\pm7.3$ &--\\
iPTF15dmu&      SN? & 0.118&      552&    $-1.036\pm0.084$ &     $6.54\pm0.70$   &-- & $41.5\pm4.1$ & $\sim 10$\\
iPTF15dni&   AGN? & 0.0191&      83&    $-1.01\pm 0.16$ &       $0.452\pm0.093$  &--   & $2.87\pm0.59$ &--\\
iPTF15fgl&     SN Ibn &  0.175&      849&    $-0.82\pm 0.25$ & $1.50\pm0.58$&    --&$9.5\pm3.7$   &    --\\
iPTF15fhl&    SN Ib/c &0.0437&      195&   $-1.01\pm0.17$& $0.63\pm0.14$   &--   & $4.00\pm0.91$ & $\sim 0.1$\\
\hline
\end{tabular}
\end{footnotesize}\end{center}\end{table*}

For and order-of-magnitude comparison, we also estimated optical SFRs for some of the iPTF transients. We fitted the public SDSS and WISE photometry using our own custom SED-fitting software, employing the \citet{bc03} population synthesis templates at solar metallicity and Calzetti dust \citep[$A_V$ free, and with $R_V$ fixed to the standard Calzetti value of 4; see][]{cal00}.  We assumed a maximum population age of 10\,Gyr and a constant star-formation history except for a transition between 10-100 Myr, which allows us to fit the young population (SFR) and old population (total mass) independently.  The resulting SFR estimates, reported in the last column of Table  \ref{tb:_SFR}, depend significantly on these assumptions but form a reasonable estimate of the average SFR over the last $\sim$10 Myr. We note that the host galaxies of iPTF15dmn and iPTF15fgl are not in the SDSS footprint so we cannot calculate their optical SFR with this method. The host galaxy of iPTF15dni shows AGN signatures in its optical spectrum so we do not estimate its optical SFR (see Section \ref{optical_propt} for discussion).

The radio spectral indices measured using our broad-band observations of the locations of iPTF15dld, iPTF15dmu, and iPTF15fgl range in between $\approx -1.2$ and $\approx -0.8$ (Table \ref{tb:_SFR}), and are consistent within the errors with the spectral indices of SF galaxies between 1.4-4.8 GHz, which are estimated to be  $-1.1\textless \alpha \textless-0.4$ \citep{sdm08}.  We note, for comparison, that a typical radio quasar spectral index is $\approx -0.5$, although $\approx -0.8$ could be more typical for quasars without the jet aligned along our line
of sight \citep[e.g.,][]{sab12}.  

\subsection{Mid-IR properties}
As a way to test further strategies for differentiating between AGN and normal galaxies, we also searched for \textit {WISE} colors information \citep[e.g.,][]{mac12} for the host galaxies of the iPTF sources with nearby radio detections. The  completeness  and  reliability  of  the  \textit{WISE} color AGN-selection technique are estimated to be $\approx 78\%$ and $\approx 95\%$, respectively \citep{sab12}.
Generally, normal radio galaxies emit a black body spectrum peaking at $1.6 \mu$m. 
Dust in AGN gain high temperatures and radiate at the mid-IR wavelengths, following a power law spectrum.
Therefore, AGN are generally expected to be much redder than normal galaxies at the mid--IR wavelengths \citep{lss04,seg05,ghj14}.  

In Table~\ref{tb:_col_col} we list the  3.4\,$\mu$m, 4.6\,$\mu$m and12\,$\mu$m Vega magnitudes (referred to as W1, W2, W3, respectively) and offsets of the \textit{WISE} sources closest to the significant radio excesses detected near the followed-up  iPTF sources.  
In Figure~\ref{wise_col_col} we plot W2-W3 vs W1-W2, and the AGN wedge defined as:
\begin{eqnarray}
\nonumber W2-W3>2.517~~~~~~~~~~~\\
\nonumber W1-W2>0.315\times\left(\rm{W2-W3}\right)-0.222
\end{eqnarray}
As evident from this Figure, all of the host galaxies of the iPTF transients followed with the VLA are clearly within the region of normal galaxies \citep{mac12,ghj14}. 

\begin{table*}
\begin{center}\begin{footnotesize}
\caption{Vega magnitudes at 3.4\,$\mu$m, 4.6\,$\mu$m, and 12\,$\mu$m of \textit{WISE} sources located close to the iPTF transients. 
The offset column lists the angular separation between the iPTF position and the \textit{WISE} source position (for the iPTF candidates). Only sources with a signal-to-noise ratio (snr) $\textgreater$2 at wavelengths corresponding to W1, W2, and W3 are listed. \label{tb:_col_col}}
\begin{tabular}{lcccllll}
\hline
\hline
name & W1 &snrw1&W2&snrw2&W3&snrw3& offset\\ & (mag)& &(mag) &&(mag) &&(arcsec)\\
\hline
iPTF15dkv &      11.892$\pm$    0.027&      40&      11.331$\pm$    0.023&      46&      7.231$\pm$    0.018&      59&      4.1\\
iPTF15dld&      14.756$\pm$    0.034&      32&      14.511$\pm$    0.068&      16&      11.27$\pm$     0.32&      3.4&      5.3\\
iPTF15dmn&      12.003$\pm$    0.025&      43&      11.584$\pm$    0.024&      46&      8.042$\pm$    0.024&      44&     0.68\\
iPTF15dmu&      13.767$\pm$    0.025&      44&      13.433$\pm$    0.032&      33&      9.441$\pm$    0.034&      32&      1.1\\
iPTF15dni&      10.652$\pm$    0.022&      49&      10.590$\pm$    0.020&      55&      7.472$\pm$    0.018&      61&     0.86\\
iPTF15fgl&      15.128$\pm$    0.034&      32&      14.771$\pm$    0.058&      19&      10.826$\pm$    0.088&      12&      1.5\\
iPTF15fhl&      15.010$\pm$    0.053&      21&      14.766$\pm$    0.084&      13&      12.456$\pm$     0.466&      2.3&      5.7\\
\hline
\end{tabular}
\end{footnotesize}\end{center}\end{table*}

\subsection{Optical properties}
\label{optical_propt}
Our iPTF follow-up observations point toward an AGN nature of the transients iPTF15dmn and iPTF15dni \citep[see][]{k16}. The presence of an AGN in iPTF15dmn is suggested by the optical variability and nuclear origin observed during our iPTF follow--up. A pre-outburst optical spectrum of iPTF15dmn host by the 6dF galaxy redshift survey \citep{bpt81,jsc+04} reveals broad H$\alpha$ emission, and a large [\ion{N}{2}] to H$\alpha$ emission line ratio in the narrow lines galaxy emission component, that also point toward the presence of an AGN in this galaxy \citep{bpt81}. Optical spectroscopy of iPTF15dni obtained \textit{after} the GW trigger reveals a similar pattern of broad H$\alpha$ emission and strong narrow emission lines of [\ion{N}{2}] and \ion{O}{3} \citep{k16}.  The spectrum is contaminated by significantly stronger stellar continuum; however, we consider it quite likely that an AGN is present in this case as well.

\begin{figure}
\begin{center}
\includegraphics[width=6.5cm,angle=-90]{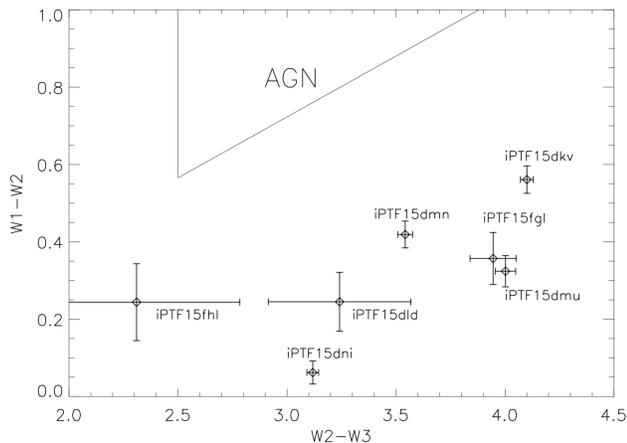}
\caption{\textit{WISE} color-color diagrams (W1-W2 vs W2-W3) of the host galaxies of the iPTF transients with a nearby radio detection (see Table \ref{tb:_col_col}).  The AGN wedge (W2-W3$\textgreater$2.517, W1-W2$\textgreater 0.315\times(W2-W3)-0.222$ \citep{mac12} is marked by solid lines. \label{wise_col_col}}
\end{center}
\end{figure}

It is interesting to note that neither iPTF15dni nor iPTF15dmn appear in the locus of AGN based on their radio emission or \textit{WISE} colors (Figs. \ref{power_vs_z} and \ref{wise_col_col}).  The fact that iPTF15dni and iPTF15dmu are not classified as AGN for their radio emission is not surprising. Indeed, \citet{sab12} have shown that because deep radio data detect emission related to stellar processes (e.g., supernova remnants) as well as AGN activity, $\approx 42\%$ of \textit{WISE}-selected AGN candidates are expected to have radio matches. However, only $\approx 2\%$ of \textit{WISE}-selected AGN with spectroscopic redshifts are radio loud according to the criteria shown in Figure \ref{power_vs_z}. On the other hand,  one possibility (suggested by our spectra of iPTF15dni) for the lack on \textit{WISE} color indication of AGN activity in iPTF15dni and iPTF15dmn is that the AGN emission is relatively weak (compared to the stellar continuum), and so accretion onto the central super-massive BH may not significantly impact the infrared colors of such galaxies.  Regardless of the actual reason for this observed phenomenon, it is clear that \textit{WISE} color or radio brightness cuts cannot be used exclusively to \textit{rule out }an AGN origin for future radio transient follow-up. 

\section{Radio sources in the \textit{Fermi}/LAT transient error circle}
\label{Sec:Fermi}

Our RMS sensitivity in the VLA follow-up observations of the \textit{Fermi}/LAT transient identified in the error region of G194575 (see Sections \ref{VLA_program}-\ref{DA}) was $\approx0.12-0.89$\,mJy at $\approx 3$\,GHz during the first epoch (which was used to extract sources; Fig.~\ref{fig:rms_vs_radius}). Close proximity to the Sun, RFI in S-band, and short integration times ($\approx 5$\,min), made our sensitivity much worse than the $\sim 10 \mu$Jy sensitivity reached in the VLA follow-up of iPTF transients.
For comparison, the RMS of the VLA FIRST survey was $\approx 0.13$ mJy at 1.4 GHz \citep[][]{bwh95} and thus, unsurprisingly, almost all radio sources found in the fields (see Table~\ref{tb:fermi}) are also previously cataloged radio sources \citep{bwh95}. 

Three radio sources identified in the error region of the possible \textit{Fermi}/LAT transient (Fermi 6, 12, and 17) do not have a clear match (i.e., a match within the estimated position errors) in the NASA/IPAC Extragalactic Database (NED)\footnote{The NASA/IPAC Extragalactic Database (NED) is operated by the Jet Propulsion Laboratory, California Institute of Technology, under contract with the National Aeronautics and Space Administration.}. These sources all have 3 GHz peak fluxes $\lesssim 1.5$\,mJy, that imply a 1.4\,GHz flux below the $3\sigma$ detection limit of VLA FIRST if we assume a self-absorbed spectral index of $\approx 2$. None of these sources show any significant variation in their radio fluxes over the timescales of our follow-up.

\section{Conclusion and future prospects}
\label{conclusion}
During O1 we have demonstrated our ability in carrying out a well-coordinated multi-band follow-up of Advanced LIGO triggers. We have followed-up 16 iPTF optical transients out of which 7 showed a significant nearby radio excess associated with host galaxy radio emission. There were two stripped envelope (Type Ib/c) SNe, and one Type Ibn SN among the iPTF transients with a nearby radio detection.  Based on the absence of significant variability in the detected radio fluxes, on the radio spectral indices, and on the SFRs derived by combining our VLA data with the redshift information provided by optical observations, we found that the all of the detected radio excesses are consistent with star forming regions or AGN activity, as also confirmed by \textit{WISE} colors. We have tested our capabilities for the follow-up of the error regions of \textit{Fermi}/LAT candidates that are too close to the Sun to be accessible to optical or X-ray facilities. In the future, this strategy may also be useful to detect radio counterparts to high energy transients hidden in dusty environments which may not be detectable via optical observations. Multiple sources were detected within the region that was observed, most of which were previously cataloged radio sources. None of these sources showed any variability between the two epochs of our observations.

\begin{figure}
\begin{center}
\includegraphics[width=8cm]{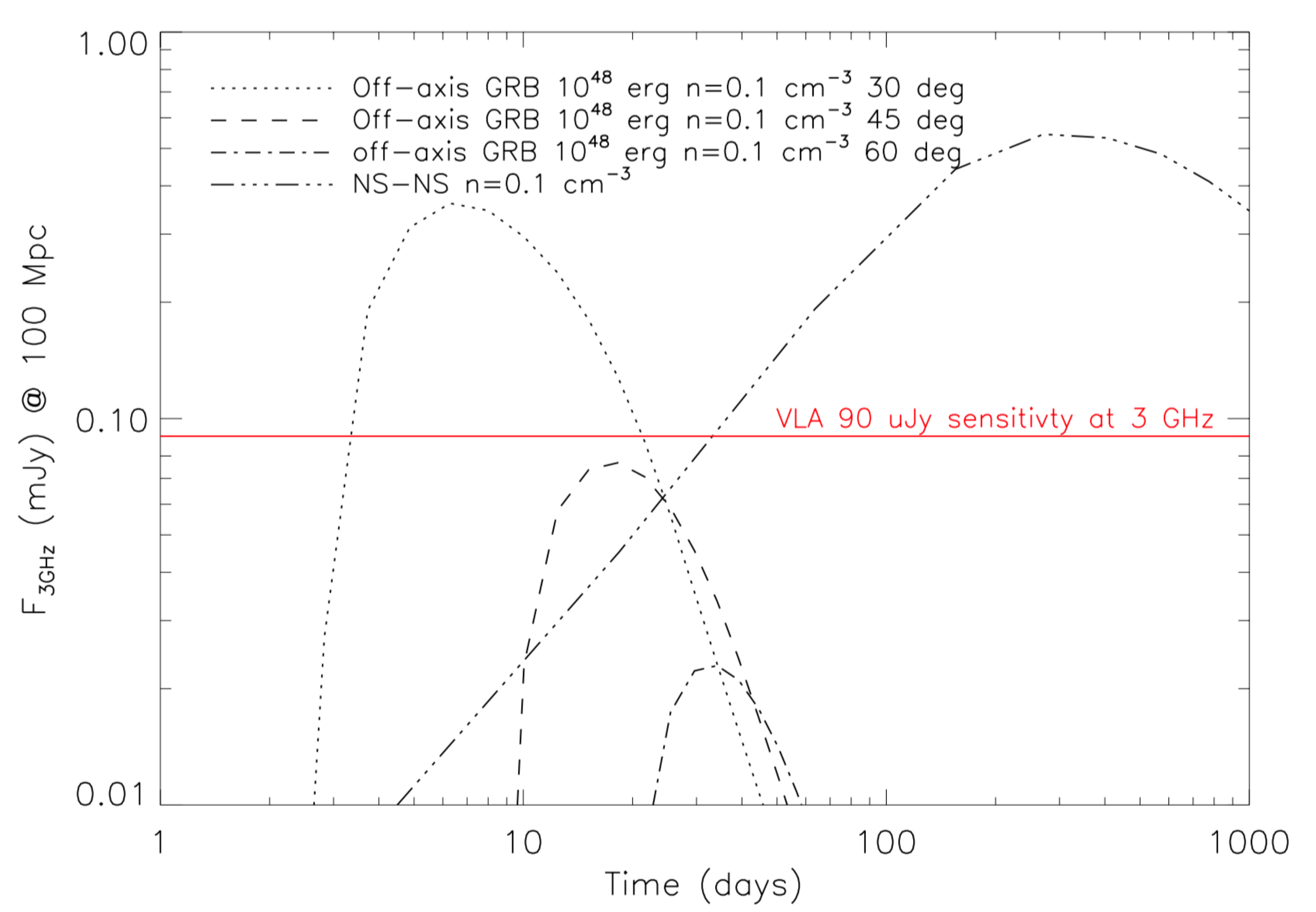}
\caption{
Model radio (3 GHz) light curves at 100\,Mpc (the expected Advanced LIGO O2 horizon distance) for a NS-NS merger with canonical slow ($E_{\rm kin}\approx 3\times10^{50}$\,erg and average $\beta c\approx 0.25$) and fast jet (off-axis GRB with $E_{\rm kin}\approx10^{48}$\,erg, $\beta c \approx 1$ at different observing angles) components. These light curves are from \citet{hnh16}. The CSM density is set to 0.1\,$\rm {cm^{-3}}$.
The horizontal red line marks the approximate $3\sigma$ VLA sensitivity at 3\,GHz reached in our follow-up. 
\label{fig:lightcurve}}
\end{center}
\end{figure}

The second observing run (O2) of Advanced LIGO is scheduled to  start in 2016, and expected to reach horizon distances of $\approx 80-120$ Mpc for binary NS mergers \citep{abbot2016g}.  
For O2, the hope is to detect at least one compact binary with a NS. 
Advanced Virgo should also come online towards the end of O2, and with a third detector the error areas of NS-NS mergers should be reduced to 200 $\rm deg^2$ \citep{spf14}.  
For O2, our team has a VLA program already in place to follow--up iPTF-identified transients in Advanced LIGO error regions. (We also have an approved \textit{Swift/XRT} GI program to follow-up the five most promising iPTF candidates, with observations that will go deeper than the NASA-led \textit{Swift/XRT} program aimed at tiling the LIGO error regions; see \citet{ekp16}). 

Recent studies \citep{chr16,hnh16} have investigated the likelihood of a radio detection from NS-NS and BH-NS mergers, and find that a large fraction of binary NS mergers occurring in realistic density environments are detectable with the VLA (Fig.~\ref{fig:lightcurve}). AGN variability, which may take place over days--year long timescales \citep[][]{mhb16,mfo13,vtu92}, could represent a source of false positives for the radio follow-up of GWs. However, as we have demonstrated in this study, combining radio observations with optical and mid-IR ones will likely identify AGN-associated variability. The VLA Sky Survey will also provide soon crucial information about the variability of the radio sky in the GHz range at the  $\approx$120 $\mu$Jy level \footnote{https://science.nrao.edu/science/surveys/vlass}. 

With plans for commissioning of the Zwicky Transient Facility in 2017 with a 47\,deg$^2$ camera, future optical transient searches will be faster and more sensitive \citep{kcs16}. Moreover, with KAGRA and LIGO India expected to join the global network in the future, the GW error areas of NS-NS systems will be reduced to tens of deg$^2$, and EM follow-ups will become much easier \citep{abbot2016g,ams13,nsd11}.  As Advanced LIGO reaches its nominal sensitivity, GW emission from core-collapse events may also come into reach  \citep{fhh02},  
although given the complexity of core-collapse progenitor models, equations of state, and explosion physics, the expected rates for these type of signals are highly uncertain.   

EM detections of GWs will provide orders of magnitude better localization, increase the confidence in low significance GW detections, and help constrain source parameters and populations properties.  In summary,  the science returns of the radio follow-up of GWs are expected to be immense. With the iPTF/ZTF and the VLA working in coordination with the advanced ground-based GW detectors, exciting discoveries are soon to revolutionize our view of the transient sky.

\begin{table*}
\begin{center}
\begin{footnotesize}
\caption{Field number, source number, VLA observation epoch, time between the \textit{Fermi} discovery and the VLA observation, VLA flux at 3\,GHz, VLA position, distance from VLA image center, VLA position error, offset between our VLA position and the closest source in NED, position error of the closest source in NED (obtained by adding in quadrature the semi-axes of the error ellipse), and the object type of the closest source as reported in NED (RadioS for a previously cataloged Radio Source; GGroup for a previously cataloged Group of Galaxies; and UvS for a previously cataloged Ultraviolet Source). The flux listed on the second epoch is measured as the maximum flux at the source position obtained from the images of the first epoch. All data were taken with the VLA in its D configuration. \label{tb:fermi}}
\begin{tabular}{clcllllllll}
\hline
\hline
Field & source n. & VLA epoch & $\Delta T$ & VLA flux & RA Dec & Cen. dist. & VLA pos. err.&NED offset & NED pos. err. & Type\\
        &                 &  (MJD)           & (day) &  (mJy beam$^{-1}$)&(hh:mm:ss~~dd:mm:ss) & (arcmin) & (arcsec)&(arcsec)&(arcsec)&\\
\hline
LAT0 &Fermi 1 &   57331.9 & 14 & $71.5\pm3.6$ & 14:46:20.366~~-03:18:08.46&    6.9 & 0.18 &      2.7&  1.9 & RadioS\\
"        &      "       &    57339.8 & 22&  $70.9\pm3.6$ & --                                            &    --    &    --    &          --  &   -- &        "      \\
"        &Fermi 2  &    57331.9 & 14 & $48.6\pm2.4$ & 14:46:53.941~~-03:11:27.01  &   4.3 &  0.14  &     1.5&   1.9 & RadioS\\
"        &       "      &    57339.8 & 22 & $37.5\pm2.0$ &          --                                          &--&--   &       --  &      --  & " \\
"        &Fermi  3 &    57331.9 & 14 & $12.28\pm0.80$ & 14:46:14.509~~-03:19:28.80&    8.8 & 1.5 &     1.3&   2.1 & RadioS\\
"        & "            &    57339.8 & 22 & $13.5\pm1.5$     &         --                                    &--&--      &     --    &       -- &          "   \\
"        &Fermi 4  &    57331.9 & 14 & $3.79\pm0.28$   & 14:46:57.834~~-03:14:08.05&     3.3 & 1.2 &    8.8& 13 & RadioS\\
"        &     "        &    57339.8&  22 & $3.39\pm0.53$   &         ---                                 &--&--         &    --      & "\\
"        &Fermi 5  &    57331.9& 14 & $10.25\pm0.78$  & 14:46:15.028~~-03:20:46.17&    9.4 & 1.2 & 0.9 &  3.2 & RadioS\\
"        &     "        &    57339.8&  22 & $14.0\pm1.7$     &           --                                &--&--           &  --        & --   & " \\
"        &Fermi 6  &    57331.9 & 14 & $1.51\pm0.21$ & 14:46:40.350~~-03:16:27.08     &   1.8 &   2.6&      39&   3.1&UvS \\
"        & "            &    57339.8&  22 & $0.84\pm0.51$  &      --                         &--&--                        &       --          & -- & "\\
"         &Fermi 7 &    57331.9& 14 & $1.55\pm0.22$   &14:46:44.840~~-03:17:58.07  &     2.7 &  3.3 & 12&    3.1 & RadioS\\
"        &  "           &    57339.8& 22 & $0.80\pm0.46$   &       --                                        & --        & --  & -- & -- & "\\
\hline
LAT1&Fermi 8  &   57331.9& 14 & $37.6\pm2.0$ & 14:47:56.718~~-02:58:18.61&    9.0& 0.50&     1.4&  1.9 & RadioS\\
"       &   "          &    57339.8& 22 &$29.9\pm1.6$ & --                                           &     --& --&     --  &         -- & "\\
"        &Fermi 9&  57331.9& 14 & $5.56\pm0.54$ &  14:47:48.325~~-02:57:48.09&   7.9 &    1.8&       2.5& 4.8 & RadioS\\
"        & "              & 57339.8&  22&$5.79\pm0.57$ & -- & --& -- &-- & -- & "\\
\hline
LAT2&Fermi 10& 57331.9 & 14 &$12.68\pm0.65$ & 14:45:43.304~~-03:05:58.19&  5.2 & 0.33& 1.5& 3.3 &  RadioS\\
 "       &   "         & 57339.8&  22&$14.22\pm0.86$ &     --                                      &--&--      &      --  &    -- & " \\
"        &Fermi 11& 57331.9&  14 &$8.50\pm0.45$  & 14:45:57.216~~-03:03:37.72&  1.3 & 0.35 &  3.3 & 4.5 &RadioS\\
"        &   "          & 57339.8&  22&$8.57\pm0.55$ &                  --                          &--&--         &       --  &    -- & "\\
"        &Fermi 12& 57331.9 & 14 &$1.16\pm0.19$ & 14:46:17.400~~-03:04:34.76&      3.8  & 3.1 &34  &      2.3&       UvS\\
"        &   "         & 57339.8   & 22&$1.47\pm0.42$ &         --                      &--&--                 & --         & -- & "\\
\hline
LAT3&Fermi 13& 57331.9 &14 & $5.33\pm0.41$ & 14:47:18.185~~-03:33:50.68&  8.7 &    1.9&          5.0 &3.8 &RadioS\\
"       &   "           & 57339.8& 22& $6.02\pm0.65$ & --                                &--&--                    &         --   &  --           & " \\
"       &Fermi 14& 57331.9 & 14 &$4.28\pm0.38$ & 14:47:29.220~~-03:16:48.32 &   8.8 &   2.7&  4.5&7.8 & RadioS\\
"       &"              & 57339.8& 22&$2.71\pm0.59$ &--                &--&--     & -- & "\\
"       &Fermi 15& 57331.9& 14 &$3.43\pm0.35$ & 14:47:47.990~~-03:32:39.87&    8.6 &  2.7 &       2.9 & 8.4&RadioS\\
"       &    "         & 57339.8& 22&$2.01\pm0.56$ &       --                       & -- &--                        & --      &  -- & "\\
"       &Fermi 16& 57331.9& 14 &$4.70\pm0.52$ & 14:46:57.650~~-03:31:56.42&  10 &     3.5&      13&  7.1 & GGroup\\
"       &               & 57339.8& 22&$7.70\pm0.95$ & --      &--&--                                               & --        & --  & "\\
"       &Fermi 17& 57331.9& 14 &$1.33\pm0.16$ & 14:47:13.630~~-03:27:54.48&     4.5 &  4.5&   --   &-- & --\\
"       &   "          & 57339.8&  22&$1.09\pm0.26$ & --  & -- & --& -- & -- & --\\ 
"       &Fermi 18& 57331.9& 14 &$0.91\pm0.14$ & 14:47:30.900~~-03:28:05.80&   2.6   &  3.2&   8.6   &1.8 & UvS\\
"       &   "          & 57339.8 & 22&$0.59\pm0.22$     &--                                       &--&--          & --        & -- & "\\
\hline
LAT4& Fermi 19& 57331.9&  14 &$201\pm10$ & 14:45:42.387~~-03:30:00.11&   6.7 &  0.22&      1.5& 2.1 & RadioS\\
"        & "            & 57339.8&  22&$227\pm11$ &  --                       &--&--                           &     --      &  --      & "\\
"        &Fermi  20&57331.9&  14 &$7.99\pm0.99$ &14:46:00.870~~-03:24:20.31&     1.3 & 6.1&      1.7 & 2.8 &RadioS\\
"        & "             &57339.8&   22&$8.98\pm0.81$ & --   &--&--                                         & --         & --     & "\\
("           & Fermi 1 & 57331.9 & 14 &$60.3\pm3.8$ &  14:46:20.592~~-03:18:05.71 & 8.7 & 0.76&  2.3 &  1.9& RadioS)\\
("           & Fermi 3 & 57331.9 & 14 &$12.4\pm1.7$ & 14:46:14.750~~-03:19:27.39 & 6.8  & 3.1 & 2.8 & 2.1 & RadioS)\\
("           & Fermi 5 & 57331.9 & 14 &$12.2\pm1.6$ & 14:46:15.050~~~-03:20:44.73 & 5.8 & 3.0 & 1.1 & 3.2 & RadioS)\\
\hline
\end{tabular}%
\end{footnotesize}\end{center}\end{table*}

\acknowledgements
 A.C. thanks K. Hotokezaka and S. Nissanke for graciously providing the theoretical radio light curves of NS-NS mergers. A.C. acknowledges support from the NSF CAREER award \#1455090. A.C. and N.P. acknowledge partial support from NASA/Swift Cycle 11 GI via grant NNX16AC12G. M.M.K. acknowledges partial support from the GROWTH project funded by the NSF under Grant \#1545949. N.M. acknowledges support from the TTU Clark Scholars program. AG-Y acknowledges support from the European Union FP7 programme through ERC grant \# 307260, 
the Quantum Universe I-Core program by the Israeli Committee for Planning and Budgeting and the ISF; by Minerva and ISF grants; and by Kimmel and YeS awards. The National Radio Astronomy Observatory is a facility of the National Science Foundation operated under cooperative agreement by Associated Universities, Inc. The Intermediate Palomar Transient Factory project is a scientific collaboration among the California Institute of Technology, Los Alamos National Laboratory, the University of Wisconsin, Milwaukee, the Oskar Klein Center, the Weizmann Institute of Science, the TANGO Program of the University System of Taiwan, and the Kavli Institute for the Physics and Mathematics of the Universe. This research used resources of the National Energy Research Scientific Computing Center, a DOE Office of Science User Facility supported by the Office of Science of the U.S. Department of Energy under Contract No. DE-AC02-05CH11231.


\begin{thebibliography}{}
\expandafter\ifx\csname natexlab\endcsname\relax\def\natexlab#1{#1}\fi

\bibitem[Abadie et al.(2010)]{abadie2010} Abadie, J., Abbott, B.~P., Abbott, R., et al.\ 2010, Classical and Quantum Gravity, 27, 173001 

\bibitem[Abbott et al.(2016a)]{abbot2016a} Abbott, B.~P., Abbott,  R., Abbott, T.~D., et al.\ 2016a, \prl, 116, 131103

\bibitem[Abbott et al. (2016b)]{abbot2016b} Abbott, B.~P., Abbott,  R., Abbott, T.~D., et al.\ 2016b, \prl , 116, 061102 

\bibitem[Abbott et al. (2016c)]{abbot2016c} Abbott, B.~P., Abbott, R., Abbott, T.~D., et al.\ 2016c,  \apjl, 826, 13 

\bibitem[Abbott et al.(2016d)]{abbot2016d} Abbott, B.~P., Abbott, R., Abbott, T.~D., et al.\ 2016d, \prl, 116, 241103 

\bibitem[Abbott et al.(2016e)]{abbot2016e} Abbott, B.~P., Abbott, R., Abbott, T.~D., et al.\ 2016e, \prl, 116, 241102 

\bibitem[Abbott et al.(2016f)]{abbot2016f}  Abbott, B.~P., Abbott, R., Abbott, T.~D., et al.\ 2016f, eprint arXiv:1606.04856

\bibitem[Abbott et al.(2016g)]{abbot2016g} Abbott, B.~P., Abbott, R., Abbott, T.~D., et al.\ 2016f, Living Reviews in Relativity, 19,  1

\bibitem[Adriani et al.(2016)]{aaa16} Adriani, O., Akaike, Y., Asano, K., et al.\ 2016, arXiv:1607.00233 

\bibitem[Aso et al.(2013)]{ams13} Aso, Y., Michimura, Y., Somiya, K., et al.\ 2013, \prd, 88, 043007 

\bibitem[Baldwin et al.(1981)]{bpt81} Baldwin, J.~A., Phillips, M.~M., \& Terlevich, R.\ 1981, \pasp, 93, 5

\bibitem[Barnes \& Kasen (2013)]{bk13} Barned, \& J., Kasen, 2013, \apj, 775, 18 

\bibitem[Becker et al.(1995)]{bwh95} Becker, R.~H., White, R.~L., \& Helfand, D.~J.\ 1995, \apj, 450, 559 

\bibitem[Berger(2010)]{b10} Berger, E.\ 2010, \apj, 722, 1946  

\bibitem[Berger(2014)]{b14} Berger, E.\ 2014, \araa, 52, 43 

\bibitem[Blandford
\& Znajek (1977)]{bz77} Blandford, R.~D., \& Znajek, R.~L.\ 1977, MNRAS, 179, 433 

\bibitem[Bowman et al. (2013)]{bck13} Bowman, J.~D., Cairns,
I., Kaplan, D.~L., et al.\ 2013, PASA, 30, e031

\bibitem[Bruzual \& Charlot (2003)]{bc03} Bruzual, G.; \& Charlot, S. 2003, \mnras, 344, 1000

\bibitem[Calzetti et al. (2000)]{cal00}	
	Calzetti, D., Armus, L., Bohlin, R.~C., Kinney, A.~L., Koornneef, J., Storchi-Bergmann, T. 2000, \apj, 533, 682

\bibitem[Cavalier et al. (2006)]{cbb06} Cavalier, F., Barsuglia, M., Bizouard, M.-A., et al.\ 2006, \prd, 74, 082004

\bibitem[Cenko et al.(2015)]{gcn18762} Cenko, S. B.\ 2015, GRB Coordinates Network, 18762, 1

\bibitem[Chu et al.(2016)]{chr16} Chu, Q., Howell, E.~J., Rowlinson, A., et al.\ 2016, \mnras, 459, 121 

\bibitem[Condon (1992)]{Condon1992} Condon, J. J. 1992, ARA\&A, 30, 575

\bibitem[Condon et al.(1998)]{ccg98} Condon, J.~J., Cotton, W.~D., Greisen, E.~W., et al.\ 1998, \aj, 115, 1693 

\bibitem[Connaughton et al. (2016)]{cbg16} Connaughton, V., Burns, E., Goldstein, A., et al.\ 2016, \apjl, 826, L6

\bibitem[Copperwheat et al.(2016)]{csp16} Copperwheat, C.~M., Steele, I.~A., Piascik, A.~S., et al.\ 2016, arXiv:1606.04574 

\bibitem[Corsi et al.(2014)]{cog14} Corsi, A., Ofek, E.~O.,
Gal-Yam, A., et al.\ 2014, ApJ, 782, 42

\bibitem[Corsi et al.(2016)]{cgk15} Corsi, A., Gal-Yam, A., Kulkarni, S.~R., et al.\ 2016, \apj, 830, 42

\bibitem[Cowperthwaite et al.(2016)]{cbs16} Cowperthwaite, P.~S., Berger, E., Soares-Santos, M., et al.\ 2016, arXiv:1606.04538 

\bibitem[Evans et al.(2016)]{ekp16} Evans, P.~A., Kennea, J.~A., Palmer, D.~M., et al.\ 2016, arXiv:1606.05001 

\bibitem[Fairhurst (2011)]{f11} Fairhurst, S.\ 2011, Classical and Quantum Gravity, 28, 105021 

\bibitem[Frail et al.(2000)]{fwk00} Frail, D.~A., Waxman, E.,
\& Kulkarni, S.~R.\ 2000, ApJ, 537, 191

\bibitem[Fryer et al. (2002)]{fhh02} Fryer, C.~L., Holz, D.~E., \& Hughes, S.~A.\ 2002, \apj, 565, 430 

\bibitem[G{\"u}rkan et al.(2014)]{ghj14} G{\"u}rkan, G., Hardcastle, M.~J., \& Jarvis, M.~J.\ 2014, \mnras, 438, 1149 

\bibitem[Hales et al. (2012)]{hmc12} Hales, C.~A., Murphy, T., Curran, J.~R., et al.\ 2012, \mnras, 425, 979

\bibitem[Jin et al. (2016)]{jin16} Jin, Z.-P., Hotokezaka, K., Li, X., et al. 2016, Nature Communications, in press.

\bibitem[Jones et al.(2004)]{jsc+04} Jones, D.~H., Saunders, W., Colless, M., et al.\ 2004, \mnras, 355, 747 

\bibitem[KamLAND Collaboration et al.(2016)]{ggh16} KamLAND Collaboration, Gando, A., Gando, Y., et al.\ 2016, arXiv:1606.07155 

\bibitem[Kasliwal et al. (2016)]{kcs16} Kasliwal, M.~M.,
Cenko, S.~B., Singer, L.~P., et al.\ 2016, arXiv:1602.08764

\bibitem[Kasliwal et al. (in prep.)]{k16} Kasliwal, M.~M. et al.\ in preparation

\bibitem[Klimenko et al. (2011)]{kvd11} Klimenko, S., Vedovato, G., Drago, M., et al.\ 2011, \prd, 83, 102001 

\bibitem[Lacy et al.(2004)]{lss04} Lacy, M., Storrie-Lombardi, L.~J., Sajina, A., et al.\ 2004, \apjs, 154, 166 

\bibitem[Law et al. (2009)]{lkd09} Law, N.~M., Kulkarni, S.~R., Dekany, R.~G., et al.\ 2009, \pasp, 121, 1395 

\bibitem[Li \& Paczy{\'n}ski(1998)]{lp98} Li, L.-X., \& Paczy{\'n}ski, B.\ 1998, \apjl, 507, L59 

\bibitem[LSC \& Virgo (2015a)]{gcn18626} LIGO Scientific Collaboration and Virgo\ 2015a, GRB Coordinates Network, 18626, 1 

\bibitem[LSC \& Virgo (2015b)]{gcn18442} LIGO Scientific Collaboration and Virgo\ 2015b, GRB Coordinates Network, 18442, 1

\bibitem[Loeb (2016)]{l16} Loeb, A.\ 2016, \apjl, 819, L21 

\bibitem[Mateos et al.(2012)]{mac12} Mateos, S., Alonso-Herrero, A., Carrera, F.~J., et al.\ 2012, \mnras, 426, 3271 

\bibitem[Mauch \& Sadler (2007)]{Mauch2007} Mauch, T., \&  Sadler, E.~M. \ 2007, \mnras, 375, 931 

\bibitem[Metzger \& Berger (2012)]{mb12} Metzger, B.~D., \& Berger, E.\ 2012, \apj, 746, 48

\bibitem[Metzger \& Fernandez (2014)]{mf14} Metzger, B.~D., \& Fernandez, R.\ 2014, \mnras, 441, 3444 

\bibitem[Mooley et al. (2013)]{mfo13} Mooley, K.~P., Frail, D.~A., Ofek, E.~O., et al.\ 2013, \apj, 768, 165

\bibitem[Mooley et al. (2016)]{mhb16} Mooley, K.~P., Hallinan, G., Bourke, S., et al.\ 2016, \apj, 818, 105

\bibitem[M{\"o}sta et al.(2010)]{mpr10} M{\"o}sta, P., Palenzuela, C., Rezzolla, L., et al.\ 2010, \prd, 81, 064017 

\bibitem[Murase et al. (2016)]{Murase2016} Murase, K., Kashiyama, K., M\'esz\'aros, P., et al. 2016, arXiv:1602.06938

\bibitem[Murphy et al. (2011)]{mur11} {Murphy}, E.~J. et al. 2011, \apj,737,67

\bibitem[Nakar \& Piran (2011)]{np11} Nakar, E., \& Piran, T.\ 2011, Nature, 478, 82 

\bibitem[Nissanke et al. (2011)]{nsd11} Nissanke, S., Sievers, J., Dalal, N., \& Holz, D.\ 2011, \apj, 739, 99 

\bibitem[Hotokezaka et al.(2016)]{hnh16} Hotokezaka, K., Nissanke, S., Hallinan, G., et al.\ 2016, arXiv:1605.09395 

\bibitem[Ofek et al.(2013)]{ofc13} Ofek, E.~O., Fox, D., Cenko, S.~B., et al.\ 2013, \apj, 763, 42 

\bibitem[Palliyaguru et al.(2015a)]{gcn18420} N.~T.~Palliyaguru et al.\,2015a, GRB Coordinates Network, 18420, 1

\bibitem[Palliyaguru et al.(2015b)]{gcn18474} N.~T.~Palliyaguru et al.\,2015b, GRB Coordinates Network, 18474, 1

\bibitem[Palliyaguru et al.(2015c)]{gcn18528} N.~T.~Palliyaguru et al.\,2015c, GRB Coordinates Network, 18528, 1

\bibitem[Palliyaguru et al.(2015d)]{gcn18560} N.~T.~Palliyaguru et al.\,2015d, GRB Coordinates Network, 18560, 1

\bibitem[Palliyaguru et al.(2015e)]{gcn18584} N.~T.~Palliyaguru et al.\,2015e, GRB Coordinates Network, 18584, 1

\bibitem[Palliyaguru et al.(2015f)]{gcn18780} N.~T.~Palliyaguru et al.\,2015f, GRB Coordinates Network, 18780, 1

\bibitem[Palliyaguru et al.(2016a)]{gcn18846} N.~T.~Palliyaguru et al.\,2016a, GRB Coordinates Network, 18846, 1

\bibitem[Palliyaguru et al.(2016b)]{gcn18873} N.~T.~Palliyaguru et al.\,2016b, GRB Coordinates Network, 18873, 1

\bibitem[Palliyaguru et al.(2016c)]{gcn18914} N.~T.~Palliyaguru et al.\,2016c, GRB Coordinates Network, 18914, 1

\bibitem[Perna et al. (2016)]{plg16} Perna, R., Lazzati, D., \& Giacomazzo, B.\ 2016, arXiv:1602.05140 

\bibitem[Perley et al. (2009)]{pnj09} Perley, R., Napier, P., Jackson, J., et al.\ 2009, IEEE Proceedings, 97, 1448  

\bibitem[Perley 
\& Perley (2013)]{pp13} Perley, D.~A., \& Perley, R.~A.\ 2013, \apj, 778, 172 

\bibitem[Racusin et al.(2016)]{rbg16} Racusin, J.~L., Burns, E., Goldstein, A., et al.\ 2016, arXiv:1606.04901 

\bibitem[Sadler et al. (1999)]{smj99} Sadler, E.~M., McIntyre, V.~J., Jackson, C.~A., \& Cannon, R.~D.\ 1999, \pasa, 16, 247

\bibitem[Seymour et al.(2008)]{sdm08} Seymour, N., Dwelly, T., Moss, D., et al.\ 2008, \mnras, 386, 1695 

\bibitem[Singer et al.(2014)]{spf14} Singer, L.~P., Price, L.~R., Farr, B., et al.\ 2014, \apj, 795, 105 

\bibitem[Singer et al.(2015a)]{gcn18337} Singer, L.\ 2015, GRB Coordinates Network, 18337, 1

\bibitem[Singer et al.(2015c)]{gcn18497} Singer, L.\ 2015, GRB Coordinates Network, 18497, 1

\bibitem[Smartt et al.(2016)]{scs16} Smartt, S.~J., Chambers, K.~C., Smith, K.~W., et al.\ 2016, arXiv:1606.04795 

\bibitem[Smolcic et al. (2008)]{Smolcic2008} Smolcic, V., et al. 2008, \apj, 177, 14

\bibitem[Smolcic et al. (2016)]{Smolcic2016} Smolcic, V., et al. 2016, arXiv:1603.05996

\bibitem[Stern et al.(2005)]{seg05} Stern, D., Eisenhardt, P., Gorjian, V., et al.\ 2005, \apj, 631, 163 

\bibitem[Stern et al.(2012)]{sab12} Stern, D., Assef, R.~J., Benford, D.~J., et al.\ 2012, \apj, 753, 30 

\bibitem[Vallisneri (2000)]{v00} Vallisneri, M.\ 2000, Physical Review Letters, 84, 3519 

\bibitem[van Haarlem et
al. (2013)]{vwg13} van Haarlem, M.~P., Wise, M.~W., Gunst, A.~W., et al.\ 2013, AA, 556, A2 

\bibitem[Valtaoja et al.(1992)]{vtu92} Valtaoja, E., Terasranta, H., Urpo, S., et al.\ 1992, \aap, 254, 80 

\bibitem[Vianello et al.(2015)]{gcn18458} Vianello, G.\ 2015, GRB Coordinates Network, 18458, 1

\bibitem[Weiler et al.(1986)]{wsp86} Weiler, K.~W., Sramek,
R.~A., Panagia, N., van der Hulst, J.~M.,
\& Salvati, M.\ 1986, ApJ, 301, 790

\bibitem[Woosley \& Bloom(2006)]{wb06} Woosley, S.~E., \& Bloom, J.~S.\ 2006, \araa, 44, 507

\bibitem[Woosley (2016)]{wb16} Woosley, S.~E., 2016, \apjl, 824, 10 

\bibitem[Yamazaki et al.(2016)]{yao16} Yamazaki, R., Asano, K., \& Ohira, Y.\ 2016, arXiv:1602.05050 

\end{thebibliography}
\end{document}